%% file: plb.tex
\pdfoutput=1
\newcommand*{\ATLASLATEXPATH}{./}
\documentclass[cernpreprint,texlive=2011,txfonts,UKenglish]{\ATLASLATEXPATH atlasdoc}
\usepackage{\ATLASLATEXPATH atlaspackage}
\usepackage{\ATLASLATEXPATH atlasphysics}
\usepackage{\ATLASLATEXPATH atlascover}
\usepackage{\ATLASLATEXPATH atlasbiblatex}

\usepackage{lineno,hyperref}
 
\usepackage{subfig}
\usepackage{epsfig}
\usepackage{dcolumn}%
\usepackage{bm}%
\usepackage{graphicx}
\usepackage{xspace}
\usepackage{collref}
\usepackage{url}
\usepackage{multirow}
\usepackage{rotating}

\AtlasTitle{Search for invisible decays of a Higgs boson using vector-boson fusion in $pp$ collisions at $\sqrt{s}=8$~TeV
with the ATLAS detector}

\author{The ATLAS Collaboration}

\AtlasRefCode{HIGGS-2013-16}

\PreprintIdNumber{CERN-PH-EP-2015-186}

\AtlasJournal{JHEP}

\AtlasAbstract{%
A search for a Higgs boson produced via vector-boson fusion and decaying into invisible particles is presented, using 20.3 fb$^{-1}$ of proton--proton collision data at a centre-of-mass energy of 8~TeV recorded by the ATLAS detector at the LHC.
For a Higgs boson with a mass of 125~\GeV, assuming the Standard Model production cross section, an upper bound of 0.28
is set  on the branching fraction of $H\to$~invisible at 95\% confidence level,
where the expected upper limit is 0.31. The results are interpreted in models of Higgs-portal dark matter where the branching fraction limit is converted 
into upper bounds on the dark-matter--nucleon scattering cross section as a function of the dark-matter particle mass, and compared to results from the 
direct dark-matter detection experiments. 
}

\AtlasCoverSupportingNote{VBF $H\to$~invisible Search}{http://cds.cern.ch/record/1626597}
\AtlasCoverSupportingNote{VBF $H\to$~invisible Search}{https://cds.cern.ch/record/1752861}

\AtlasCoverCommentsDeadline{19th July 2015}

\AtlasCoverAnalysisTeam{J.~Abdallah, B.~Acharya, O.~Arnaez, K.~A.~Assamagan, A.~Bassalat, R.~Bonde, P.~Calfayan, D.~Delgove, M.~Frate, E.~Lipeles, C.~Mills, A.~Nelson, W.~Quayle, R.~Rezvani, D.~Salek, C.~Sandoval, K.~Shaw, L.~Truong, R.~Vanguri, A.~White, D.~Whiteson}

\AtlasCoverEdBoardMember{Ariel Schwartzman~(chair), Pierre-Hugues Beauchemin, Naoko Kanaya, Chris Hays}

\AtlasCoverEgroupEditors{atlas-higg-2013-16-editors@cern.ch}
\AtlasCoverEgroupEdBoard{atlas-higg-2013-16-editorial-board@cern.ch}

\clearpage
 
\journal{JHEP}

\begin{document}

\maketitle

\section{Introduction}
\label{sec:intro}
Astrophysical observations provide strong evidence for dark matter (see Ref.~\cite{Clowe} and the references therein).
Dark matter (DM) may be explained by the existence of weakly interacting massive
particles (WIMP)~\cite{Goldberg:1983nd,Ellis:1983ew}. The observed Higgs boson with a mass of about 125~GeV~\cite{ATLAS:2012af,Chatrchyan:2012ia} 
might decay to dark matter or neutral long-lived massive particles~\cite{Antoniadis:2004se,ArkaniHamed:1998vp,Datta:2004jg,Kanemura:2010sh,Djouadi:2011aa},
provided this decay is kinematically allowed. This is referred to as an invisible decay of the Higgs
boson~\cite{Shrock:1982kd,Choudhury:1993hv,Eboli:2000ze,Davoudiasl:2004aj,Godbole:2003it,Ghosh:2012ep,Belanger:2013kya,Curtin:2013fra}. 

This paper presents a search for invisible decays of a Higgs boson produced via the vector-boson fusion (VBF) process. In the 
Standard Model (SM), the process $H \rightarrow ZZ \rightarrow 4\nu$ is
an invisible decay of the Higgs boson, but the branching fraction (BF) is $0.1$\%~\cite{LHCHiggsCrossSectionWorkingGroup:2011ti,LHCHiggsCrossSectionWorkingGroup:2012vm},
which is below the sensitivity of the search presented in this paper. In addition to the VBF Higgs boson signal itself, there is 
a contribution to Higgs boson production from the gluon fusion plus 2-jets (ggF+2-jets) process, which is smaller than the VBF signal in the phase space of interest in this search.
The ggF+2-jets contribution is treated as signal.
The search is performed with a dataset corresponding to an integrated luminosity of 20.3 fb$^{-1}$ of
proton--proton collisions at $\sqrt{s}=8$~TeV, recorded by the ATLAS
detector at the LHC~\cite{Aad:2008zzm}. 

The signature of this process
is two jets with a large separation in pseudorapidity\footnote{ATLAS uses a right-handed coordinate system with its origin at the nominal interaction point (IP) in the centre of the detector and the $z$-axis along the beam direction.  The $x$-axis points from the IP to the centre of the LHC ring, and the $y$-axis points upward. Cylindrical coordinates $(r,\phi)$ are used in the transverse plane, $\phi$ being the azimuthal angle around the beam direction. The pseudorapidity is defined as $\eta = -\ln \tan{\theta/2}$, where $\theta$ is the polar angle.} and large missing transverse momentum\footnote{The transverse momentum is the component of the momentum vector perpendicular to the beam axis.} $E_{{\rm T}}^{\rm miss}$.
 The VBF process, in its most extreme topology 
(high dijet invariant mass for example), offers strong rejection against the %
QCD-initiated $(W,Z)$+jets ($V$+jets) backgrounds. The resulting selection has a significantly better signal-to-background ratio than selections targeting the ggF process.

The CMS Collaboration obtained an
upper bound of 58\% on the branching fraction of invisible Higgs boson decays using
a combination of the VBF and $ZH$ production modes~\cite{Chatrchyan:2014tja}.  Weaker limits 
 were obtained using the $Z(\rightarrow \ell \ell)H + E_{{\rm T}}^{\rm miss}$ signature
by both the ATLAS and CMS collaborations~\cite{ATLAS-CONF-2013-011,Chatrchyan:2014tja},
giving upper limits at 95\% CL of 75\% and 83\% on the branching
fraction of invisible Higgs boson decays, respectively. By combining the searches in the $Z(\rightarrow \ell\ell)H$ and $Z(\rightarrow b\bar{b})H$ channels, 
CMS obtained an upper limit of 81\%~\cite{Chatrchyan:2014tja}. 
 Using the associated production with a vector boson, $VH$, where the vector-boson decays to jets and the Higgs boson to invisible particles, 
ATLAS set a 95\% CL upper bound of 78\% on the branching fraction of $H\to$~invisible~\cite{atlas:VjjHinv}.  
Other searches for large $E_{{\rm T}}^{\rm miss}$ in
association with one or more jets were reported 
in Refs.~\cite{Aad:2015zva,Khachatryan:2014rra,monoV,monojetTheory}. These searches
are less sensitive to Higgs-mediated interactions than the search presented here,
because they are primarily sensitive to the ggF process and have significantly larger 
backgrounds.

Assuming that the couplings of the Higgs boson to SM particles correspond to the SM values, global fits to measurements of cross sections 
times branching fractions of different channels allow the extraction of a limit on the Higgs boson's branching fraction to invisible particles. The 95\% CL upper limits on this 
branching fraction set by ATLAS and
CMS are $23$\% and $21$\% respectively
\cite{Aad:2015pla,Khachatryan:2014jba}. There is an important
complementarity between direct searches for invisible decays of Higgs bosons 
and indirect constraints on the sum of invisible and undetected
decays.  A simultaneous excess would confirm a signal, while a non-zero
branching fraction of $H\to$~invisible in the global fit, but no excess in the searches for Higgs boson decays to invisible particles, would
point toward other undetected decays or model assumptions as the
source of the global fit result.

In the search presented in this paper, the events observed in data are consistent with the background estimations. An upper bound on the cross section times the branching fraction of the Higgs boson decays to invisible particles is computed using a
maximum-likelihood fit to the data with the profile likelihood-ratio test statistic~\cite{2011EPJC...71.1554C,eCowan}. A constraint on the branching fraction
alone is obtained assuming the SM VBF and ggF production cross sections, acceptances and efficiencies,
for invisible decays of a Higgs boson with a mass $m_H=125$~GeV.

In the context of models where dark matter couples to the SM particles primarily through the Higgs boson \cite{higgs_portal},
limits on the branching fraction of invisible Higgs boson decays can be
interpreted in WIMP--nucleon interaction models~\cite{Fox:2011pm}
and compared to experiments which search for dark-matter particles via their direct interaction with the material of a detector~\cite{Angloher:2014myn,Agnese:2013jaa,Agnese:2014aze,Aprile:2013doa,Akerib:2013tjd,Bernabei:2008yi,Aalseth:2014jpa,Angloher:2011uu,Agnese:2013rvf}. %

The paper is organized as follows. The ATLAS detector is briefly described in Section~\ref{sec:Detector}. The modelling of the signal and background is presented in 
Section~\ref{sec:simu}. The dataset, triggers, event reconstruction, and event selection are described in Section~\ref{sec:evt}. 
The background estimations are presented in Section~\ref{sec:CR}. In Section~\ref{sec:syst}, the systematic uncertainties are discussed. The results are shown in Section~\ref{sec:results}, and model interpretations 
are given in Section~\ref{sec:interpretation}. Finally, concluding remarks are presented in Section~\ref{sec:conc}.  

\section{Detector}{
\label{sec:Detector}

ATLAS is a multipurpose detector with a forward-backward symmetric cylindrical geometry, described in detail in Ref.~\cite{Aad:2008zzm}.

At small radii from the beamline, the inner detector, immersed in a $2$~T magnetic field produced by a thin superconducting solenoid located directly inside the calorimeter, is made up of fine-granularity pixel and microstrip silicon detectors covering the range $|\eta|<2.5$, and a gas-filled straw-tube transition-radiation tracker (TRT) in the range $|\eta|<2$. The TRT complements the silicon tracker at larger radii and also provides electron identification based on transition radiation.  The electromagnetic (EM) calorimeter is a lead/liquid-argon sampling calorimeter with an accordion geometry. The EM calorimeter is divided into a barrel section covering $|\eta| <1.475 $ and two end-cap sections covering $1.375<|\eta| <3.2 $. For $|\eta| <2.5 $ it is divided into three layers in depth, which are finely segmented in $\eta$ and $\phi$.  An additional thin presampler layer, covering $|\eta| <1.8 $, is used to correct for fluctuations in energy losses between the production vertex and the calorimeter. Hadronic calorimetry in the region $|\eta| <1.7 $ uses steel absorbers, and scintillator tiles as the active medium.  Liquid-argon calorimetry with copper absorbers is used in the hadronic end-cap calorimeters, which cover the region $1.5<|\eta| <3.2 $.  A forward calorimeter using copper or tungsten absorbers with liquid argon completes the calorimeter coverage up to $|\eta| =4.9 $. The muon spectrometer (MS) measures the
curvature
of muon trajectories with $|\eta| <2.7 $, using three stations of precision drift tubes, with cathode strip chambers in the innermost layer for $|\eta| > 2.0 $. The deflection is provided by a toroidal magnetic field with an integral of approximately $3$~Tm and $6$~Tm in the central and end-cap regions of the ATLAS detector, respectively. The MS is also instrumented with dedicated trigger chambers, namely resistive-plate chambers in the barrel and thin-gap chambers in the end-cap, covering $|\eta| <2.4 $.

}

\section{Simulation}
\label{sec:simu}
Simulated signal and background event samples are produced with Monte Carlo (MC) event generators, and passed through a {\sc Geant4}\cite{GEANT4}
simulation of the ATLAS detector \cite{Aad:2008zzm,:2010wqa}, or a fast simulation based on a 
parameterization of the response to the electromagnetic and hadronic
showers in the ATLAS calorimeters~\cite{FastCaloSim} and a detailed simulation of other parts of the detector and the trigger system.
The results based on the fast simulation are validated against fully simulated samples and the difference is
found to be negligible. The simulated events are reconstructed
with the same software as the data. Additional $pp$ collisions in the
same and nearby bunch-crossings (pileup) are included by merging diffractive and non-diffractive
$pp$ collisions simulated with {\sc Pythia}-8.165~\cite{pythia81}.  The multiplicity distribution of these pileup
collisions is re-weighted to agree with the
distribution in the collision data. 

Both the VBF and ggF
signals are modelled using
{\sc Powheg-Box}~\cite{powheg1,powheg2,powheg3,powheg4,powheg5,Bagnaschi:2011tu} with CT10 parton distribution functions (PDFs)~\cite{ct10}, and {\sc Pythia}-8.165
simulating the parton shower, hadronization and underlying event\footnote{The
invisible decay of the Higgs boson is simulated by forcing the Higgs
boson (with $m_H=125$~GeV) to decay via $H\rightarrow ZZ^*\rightarrow 4\nu$.}. The VBF and ggF Higgs boson production cross sections
and their uncertainties are taken from Ref.~\cite{Heinemeyer:2013tqa}.
The transverse momentum ($p_{\rm T}$) distribution of the VBF-produced Higgs boson is re-weighted to
reflect electroweak (EW) radiative corrections computed by {\sc HAWK}-2.0~\cite{hawk}. These EW corrections amount to 10--25\% in the Higgs boson $p_{\rm T}$ range of
150--1000~GeV. The ggF contribution to the signal is re-weighted~\cite{Aad:2014eha,ATLAS:2014aga}  so that the $p_{\mathrm{T}}$ distribution of the Higgs boson in events
with two or more associated jets matches that of the next-to-leading-order (NLO) ggF+2-jets calculation in
{\sc Powheg-Box} MiNLO~\cite{Hamilton:2012np}, and the inclusive distributions in jets match that of the next-to-next-to-leading-order (NNLO) and next-to-next-to-leading-logarithm (NNLL) 
calculation in {\sc HRes}-2.1~\cite{deFlorian:2011xf,Grazzini:2013mca}.  The effects of finite quark masses are also
included~\cite{Bagnaschi:2011tu}.

\begin{figure*}[!tbhp]
\begin{center}
\subfloat[\label{feyn_signal} Signal ]{\includegraphics[height=0.105\textheight]{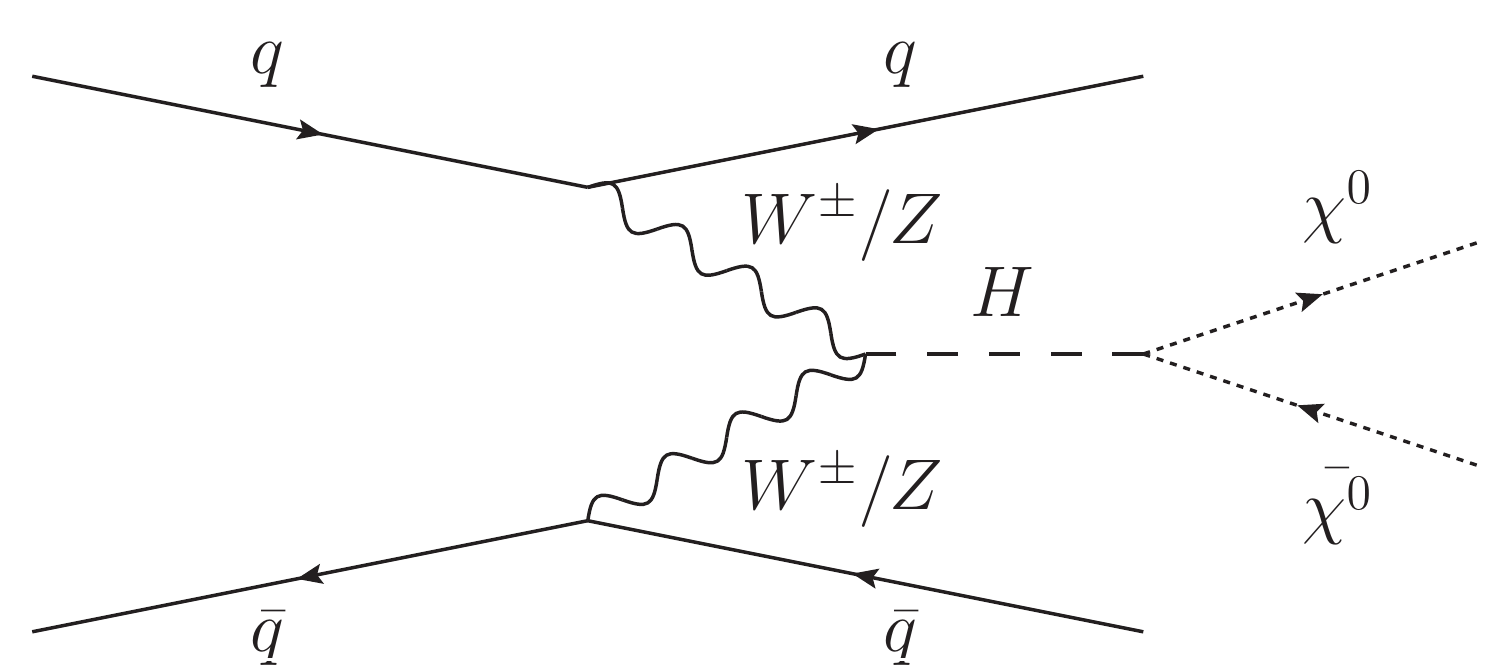}}
\subfloat[\label{feyn_strong} Strongly produced (QCD) $Z$+jets  ]{\includegraphics[height=0.105\textheight]{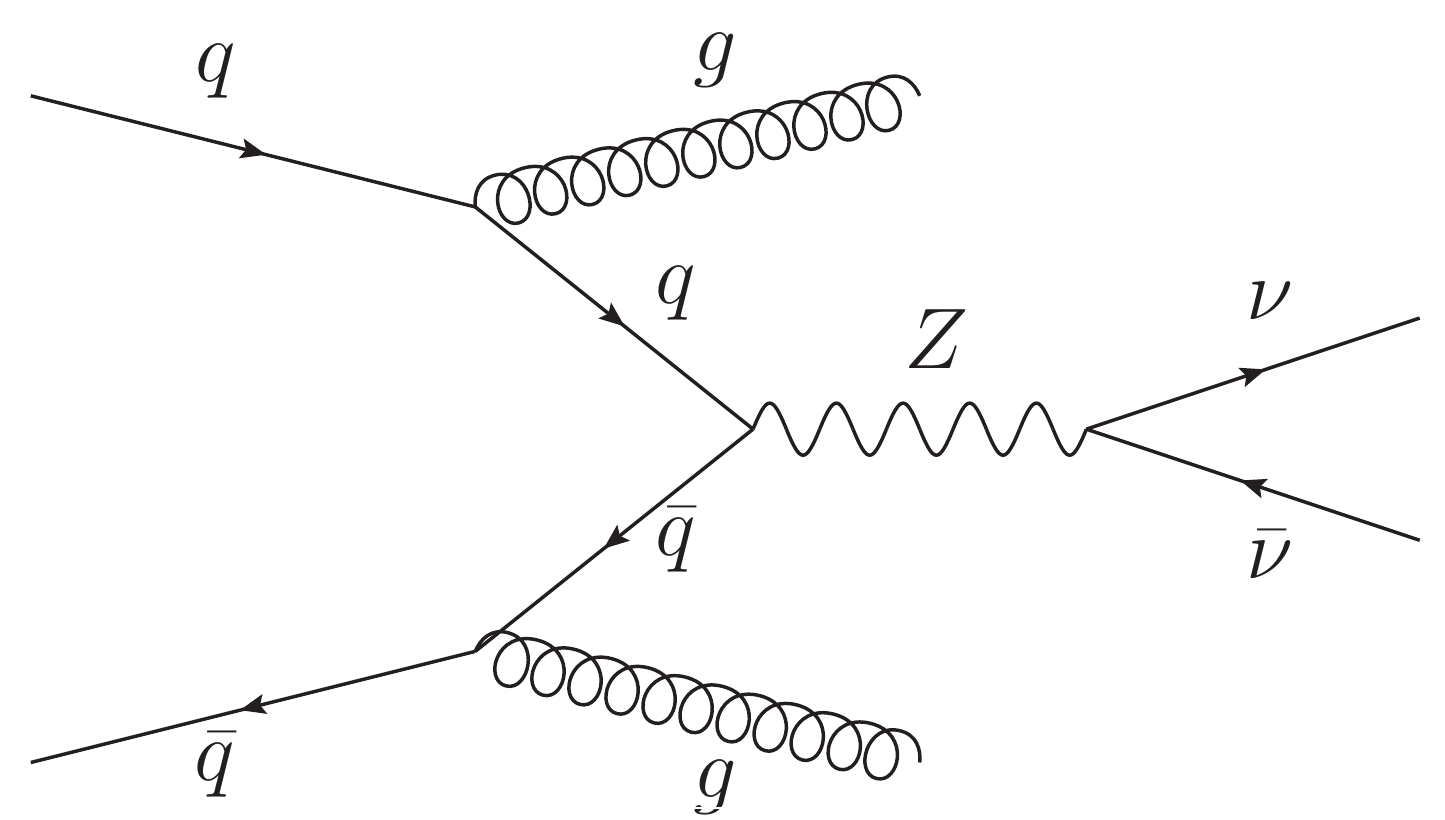}}
\subfloat[\label{feyn_weak} Weakly produced (EW) $Z$+jets  ]{\includegraphics[height=0.105\textheight]{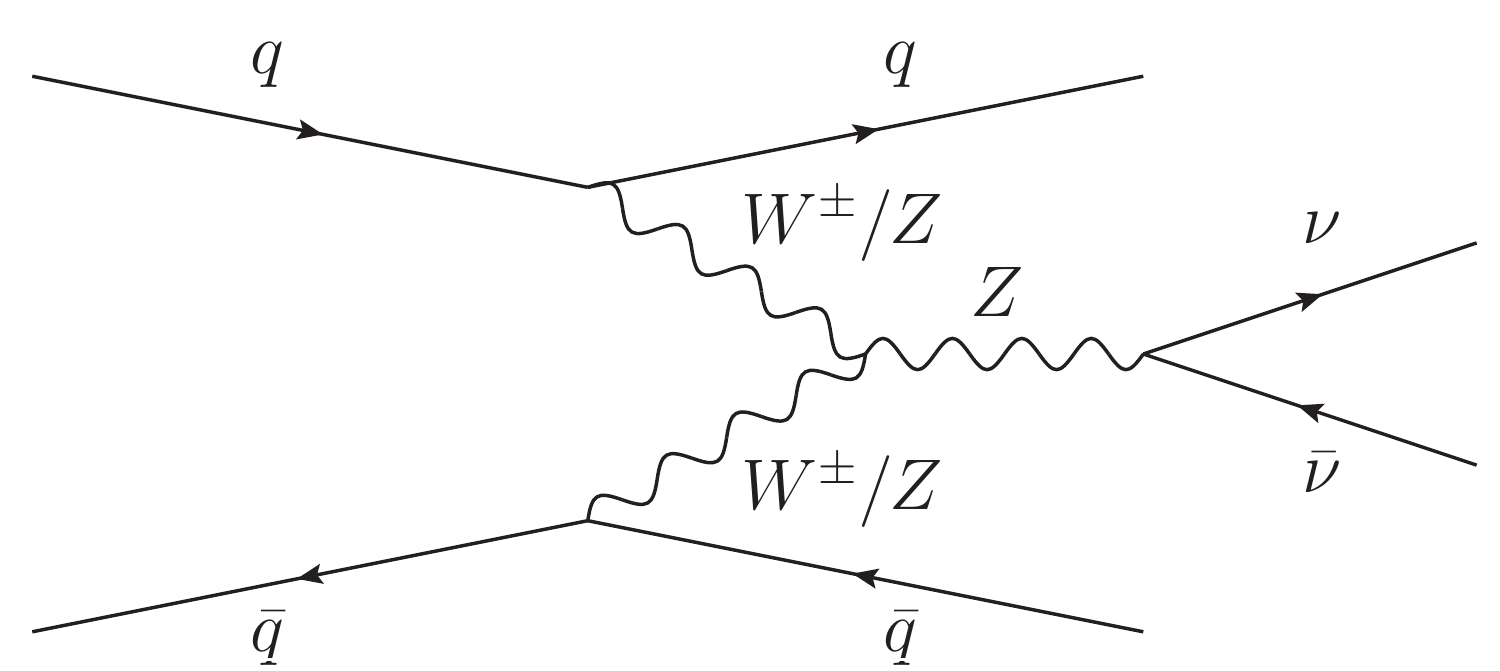}}
\end{center}
  \caption{Example Feynman diagrams for the VBF $H(\to$~invisible) signal and the vector-boson backgrounds.}
  \label{feyndiags}
\end{figure*}

The $W(\to \ell\nu)$+jets and $Z(\to\ell\ell)$+jets processes are generated using {\sc Sherpa}-1.4.5~\cite{Gleisberg:2008ta} including leading-order (LO) matrix elements 
for up to five partons in the final state with CT10 PDFs and matching these matrix elements with the parton shower following 
the procedure in Ref.~\cite{Hoche:2009}. 
The $W(\to \ell\nu)$+jets and $Z(\to\ell\ell)$+jets processes are divided into two components based on the number of electroweak vertices in the Feynman
diagrams. Diagrams which have only two electroweak vertices contain jets that are produced via the strong interaction, and are labelled ``QCD'' $Z$+jets or $W$+jets. Diagrams which have 
four electroweak vertices contain jets that are produced via the electroweak interaction, and are labelled ``EW'' $Z$+jets or $W$+jets~\cite{Aad:2014dta}.
 The MC predictions of the QCD components of $W$+jets and $Z$+jets are
normalized to NNLO in {\sc FEWZ}~\cite{Melnikov:2006kv,Anastasiou:2003ds}, while the
EW components are normalized to
VBFNLO~\cite{Arnold:2008rz}, including the jet $\pT$ and dijet invariant mass requirements.
The interference between the QCD and EW
components of $Z$+jets and $W$+jets is evaluated with {\sc Sherpa}-1.4.5 to be 7.5--18.0\% of
the size of the EW contribution depending on the signal regions. To account for this interference effect, the EW contribution 
is corrected with the estimated size of the interference term. Figure~\ref{feyndiags} shows Feynman diagrams for the signal and example vector-boson backgrounds.
There are additional small backgrounds from $\ttbar$, single top, diboson and multijet production. The $\ttbar$
process is modelled using {\sc Powheg-Box}, with {\sc Pythia}-8.165
modelling the parton shower, hadronization and underlying event.  Single-top production samples are generated with 
{\sc MC@NLO}~\cite{Frixione:2005vw} for the $s$- and $Wt$-channel~\cite{Frixione:2008yi},  while  {\sc AcerMC}-v3.8~\cite{Kersevan:2002dd} 
is used for single-top production in the $t$-channel.  A top-quark mass of 172.5~\GeV\ is  used consistently. The AUET2C (AUET2B)~\cite{ATLAStunes} set of optimized parameters for the underlying event description is used 
for $\ttbar$ (single-top) processes,  with CT10 (CTEQ6L1)~\cite{Pumplin:2002vw} PDFs. 
Diboson samples $WW$, $WZ$ and $ZZ$ (with leptonic decays) 
are normalized at NLO and generated using {\sc Herwig}-6.5.20~\cite{HERWIG} with CT10 PDFs, including
the parton shower and hadronization, and Jimmy~\cite{jimmy} to model the
underlying event, whereas the $WW$, $WZ$, and $ZZ$ ($\rightarrow \ell\ell qq,\nu\nu qq$) processes are generated together with EW $W$+jets and $Z$+jets samples. Diboson $WW$, $WZ$  and $ZZ$ ($\rightarrow \ell\ell qq,\nu\nu qq$) samples generated using {\sc Sherpa}-1.4.5 with CT10 PDFs and normalized to NLO in QCD~\cite{Campbell:2011bn} are used as a cross-check. Multijet and $\gamma$+jet samples are generated using {\sc Pythia}-8.165 with CT10  PDFs.

\section{Event selection}
\label{sec:evt}
The data used in this analysis were recorded with an $E_{\rm T}^{\rm miss}$ trigger during periods when all
ATLAS sub-detectors were operating under nominal conditions.  The trigger consists of three levels
of selections. The first two levels, L1 and L2, use as inputs coarse-spatial-granularity analog (L1) and digital (L2) sums of the measured energy.
In the final level, calibrated clusters of cell energies in the calorimeter \cite{atlaspub:topo}  are used.
At each level, an increasingly stringent threshold is applied. The most stringent requirement is  $E_{{\rm T}}^{\rm miss} >= $~80~GeV.
 Because of further corrections made in the offline reconstructed $E_{{\rm T}}^{\rm miss}$ and the resolutions
of the L1 and L2 calculations, this trigger
is not fully efficient until the offline $E_{{\rm T}}^{\rm miss}$  is greater than 150~\GeV.

Jets are reconstructed
from calibrated energy clusters\cite{atlaspaper:JES7TeV,Aad:2011he} using the 
anti-$k_{t}$ algorithm~\cite{AntiKt} with radius parameter
$R=0.4$. Jets are corrected for pileup using the event-by-event jet-area subtraction method~\cite{Cacciari:2008gn,Cacciari:2007fd} and calibrated to particle level by a 
multiplicative jet energy scale factor~\cite{atlaspaper:JES7TeV,Aad:2011he}.  The selected jets are
required to have $p_{\rm T}>20$~\GeV~and $|\eta|<4.5$. To discriminate against jets originating from minimum-bias interactions, 
selection criteria are applied to ensure that at least 50\% of the jet's summed scalar track $p_{\rm T}$, for jets within $|\eta|<2.5$, is associated with tracks originating 
from the primary vertex, which is taken to be the vertex with the highest summed $p^2_{\mathrm{T}}$ of associated tracks.
Information
about the tracks and clusters in the event is used to construct
multivariate discriminators to veto events with $b$-jets and
hadronic $\tau$-jets.  The requirements on these
discriminators identify $b$-jets with 80\% efficiency (estimated using $t\bar{t}$ events) 
\cite{ATLAS-CONF-2011-102,ATLAS-CONF-2014-004,ATLAS-CONF-2014-046}, one-track jets from hadronic $\tau$ decays with 60\% efficiency (measured with $Z\to\tau\tau$ events), and multiple-track jets from hadronic $\tau$ decays with 55\% efficiency
\cite{Aad:2014rga}. 

Electron candidates are reconstructed from clusters of energy deposits in the electromagnetic calorimeter matched to tracks in the inner detector~\cite{electronperf2010}. Muon candidates are reconstructed by requiring a match between a track in the inner detector and a track in the muon spectrometer~\cite{Aad:2014rra}.

The selection defines three orthogonal signal regions (SR), SR1, SR2a and SR2b. They are distinguished
primarily by the selection requirements on the invariant mass $m_{jj}$ of the two highest-$p_{\rm T}$ jets and their separation in pseudorapidity $\Delta\eta_{jj}$ as shown 
in Table~\ref{tab:kinematic_requirements_summary}.
\begin{table*}[tbph!]
\caption{\label{tab:selections} Summary of the main kinematic requirements in the three signal regions.}
\begin{center}
{%
\begin{tabular}{ c | c | c | c }
\hline
\hline
Requirement                                     & SR1                           & SR2a          & SR2b\\
\hline
Leading Jet $p_{\rm T}$                 & $>$75~GeV                             & $>$120~GeV            & $>$120~GeV\\
\hline
Leading Jet Charge Fraction                     & N/A                                   & $>$10\%               & $>$10\%\\
\hline
Second Jet $p_{\rm T}$                  & $>$50~GeV                             & $>$35~GeV             & $>$35~GeV\\
\hline
$m_{jj}$                                & $>$1~TeV              & $0.5<m_{jj}<1$~TeV & $>1 $~TeV \\
\hline
$\eta_{j1}\times\eta_{j2}$              &        & \multicolumn{2}{c}{$<$0}\\
\hline
$|\Delta\eta_{jj}|$                     & $>$4.8                                &  $>$3    & $3< |\Delta\eta_{jj}| < 4.8$\\
\hline
$|\Delta\phi_{jj}|$                     & $<$2.5                                & \multicolumn{2}{c}{N/A}\\
\hline
Third Jet Veto $p_{\rm T}$ Threshold            & \multicolumn{3}{c}{30~GeV}\\
\hline
$|\Delta\phi_{j,E_{\rm T}^{\rm miss}}|$ & $>$1.6 for $j_{1}$, $>$1 otherwise    & \multicolumn{2}{c}{$>$0.5}\\
\hline
$E_{\rm T}^{\rm miss}$                  & $>$150~GeV                            & \multicolumn{2}{c}{$>$200~GeV}\\
\hline
\hline
\end{tabular}
}
\label{tab:kinematic_requirements_summary}
\end{center}
\end{table*}
 The SR1 selection requires events to have two jets: one with $p_{\rm T}>75$~\GeV\ and one with $p_{\rm T}>50$~\GeV. The $\vec{E}_{{\rm T}}^{\rm miss}$ is constructed 
as the negative vectorial sum of the transverse momenta of all calibrated objects (identified electrons, muons, photons, hadronic decays of $\tau$-leptons, and jets)
and an additional term for transverse energy in the calorimeter not included in any of these objects~\cite{ATLAS-CONF-2013-082}. Events must have 
$E_{{\rm T}}^{\rm miss}>150$~\GeV~in order to suppress the background from multijet events. To further 
suppress the multijet background, the two leading jets are
required to have an azimuthal opening angle $|\Delta\phi_{jj}|<2.5$~radians and an azimuthal opening angle with respect to the $E_{\rm T}^{\rm miss}$
of $|\Delta\phi_{j,E_{\rm T}^{\rm miss}}|>1.6$~radians for the leading jet and $|\Delta\phi_{j,E_{\rm T}^{\rm miss}}|>1$ radian otherwise.
In the VBF process, the forward jets tend to have large separations in pseudorapidity ($\Delta\eta_{jj}$),
with correspondingly large dijet masses, and little hadronic activity between the two jets.
To focus on the VBF production,
the leading jets are required to be well-separated in pseudorapidity $|\Delta\eta_{jj}|>4.8$, and
have an invariant mass $m_{jj}>1$~TeV. Events are rejected if any jet is identified as
arising from the decay of a $b$-quark or a $\tau$-lepton.  The
rejection of events with $b$-quarks suppresses top-quark backgrounds.
Similarly, rejection of events with a $\tau$-lepton suppresses the $W (\rightarrow
\tau \nu)$+jets background. %
Further, events are vetoed if they contain any reconstructed leptons passing the transverse momentum  
thresholds $p_{\rm T}^e>10$~\GeV~for electrons, $p_{\rm T}^\mu>5$~\GeV~for muons, or $p_{\rm T}^\tau>20$~\GeV~for $\tau$-leptons.
Finally, events with a third jet having $p_{\rm T}>30$~\GeV~and $|\eta|<4.5$ are rejected.  
The SR2 selections are motivated by a search for new phenomena in final states with an energetic jet and large missing transverse momentum~\cite{Aad:2015zva}, and differ from those of SR1.  First, the leading jet\footnote{The ``charge fraction" of this jet is defined
as the ratio of the $\Sigma\pT$ of tracks associated to the jet to the calibrated jet $\pT$; this quantity must be at least 10\% of the maximum 
fraction of the jet energy deposited in one calorimeter layer. The charged fraction requirement  was shown to suppress fake jet backgrounds from beam-induced effects and cosmic-ray events~\cite{Aad:2015zva}.} is required to have $\pt>120$~\GeV~and $|\eta|<2.5$. Additionally, the sub-leading jet is required to have $\pT>35$~GeV, 
the $\Delta\phi_{jj}$ requirement is removed, the requirement on $\Delta\phi_{j,E_{\rm T}^{\rm miss}}$ is relaxed to $|\Delta\phi_{j,E_{\rm T}^{\rm miss}}|>0.5$, and the $\met$ requirement
is tightened to $\met>200$~GeV. A common threshold of $p_{\rm T}=7$~\GeV~is used to veto events with electrons and muons, and no $\tau$-lepton veto is applied. Finally in SR2, the $E_{{\rm T}}^{\rm miss}$ computation excludes the muon contribution and treats hadronic taus like jets (this allows the modelling of $W$+jets and $Z$+jets in the control regions  and signal regions using the same $E_{{\rm T}}^{\rm miss}$ variable as discussed in Section~\ref{sec:CR}). SR2 is further subdivided into SR2a with $500<m_{jj}<1000$~GeV, $\eta_{j1}\times\eta_{j2}<0$, and $|\Delta\eta_{jj}|>3$, and SR2b with $m_{jj}>1000$~GeV, $\eta_{j1}\times\eta_{j2}<0$ and $3<|\Delta\eta_{jj}|<4.8$.

\section{Background estimations}
\label{sec:CR} 

In order to reduce the impact of theoretical and experimental uncertainties, 
the major backgrounds, $Z\to\nu\nu$ and $W\to\ell\nu$, are determined from measurements in a set of control samples consisting of $Z\rightarrow\ell\ell$ or $W\rightarrow\ell\nu$ events ($\ell=e/\mu$).   
In each of these control regions (CR), two additional jets are required, following the same requirements as the signal regions.
The $Z\rightarrow\ell\ell$ control samples consist of events where the invariant mass of two same-flavour and opposite-sign leptons is consistent with the $Z$-boson mass, 
and so backgrounds in these control regions are small enough that they are taken from their MC predictions rather than from data-driven methods. 
 In the $W\rightarrow\ell\nu$ control regions, the background from jets misidentified as leptons is more important, at least for the case of $W\rightarrow e\nu$. In SR1, 
the background from jets misidentified as leptons in the $W\rightarrow\ell\nu$ control regions is normalized using a fit that takes advantage of the distinctive shape of 
the transverse mass distribution 
\begin{equation}
\label{eq:transmass}
m_{\rm T} = \sqrt{2  p_{\rm T}^\ell E_{\rm T}^{\rm miss} \left[1-\cos(\Delta\phi_{\ell,E_{\rm T}^{\rm miss}})\right]}  
\end{equation}
of the lepton and $E_{\rm T}^{\rm miss}$, and the charge asymmetry in $W^+/W^-$ events. In SR2, the background from jets misidentified as leptons in the $W\rightarrow\ell\nu$ 
control regions is reduced by the requirements on $m_{\rm T}$ and $E_{\rm T}^{\rm miss}$ as discussed in Section~\ref{sec:EW}. 

In order to use the control regions rather than 
the MC predictions for setting the $W$+jets and $Z$+jets background normalizations, the MC predictions in each of the three signal regions and six corresponding 
$Z(\rightarrow ee/\mu\mu)$+jets and $W(\rightarrow e\nu/\mu\nu)$+jets control regions are scaled by free parameters $k_{i}$. There is one $k_{i}$ for each signal region and the corresponding control regions. In SR1 for example, omitting factors that model systematic uncertainties, the
expected number of events for $Z(\rightarrow\nu\nu)$+jets in the signal region is
$Z_{\rm SR1}=k_{1}Z_{\rm SR1}^{\rm MC}$, for $Z(\rightarrow\ell\ell)$+jets 
in the $Z\rightarrow\ell\ell$ control region is $Z_{\rm CR}=k_{1}Z_{\rm CR}^{\rm MC}$, and for $W\rightarrow\ell\nu$ in the $W\to\ell\nu$ control region is
$W_{\rm CR}=k_{1}W_{\rm CR}^{\rm MC}$. 
 The scale factors $k_{i}$ are common for the $Z$+jets and $W$+jets background normalizations. 
The scale factors $k_i$ are determined from the maximum Likelihood fit described in Section~\ref{sec:results}.  The $Z(\to\ell\ell)$+jets 
and the $W(\to\ell\nu)$+jets MC predictions thus affect the final estimates of $Z(\to\nu\nu)$+jets and $W(\to\ell\nu)$+jets in the signal region through an implicit dependence 
on the MC ratio $Z_{\rm SR}/Z_{\rm CR}$ and  $Z_{\rm SR}/W_{\rm CR}$ for $Z(\to\nu\nu)$+jets, and   $W_{\rm SR}/W_{\rm CR}$ for $W(\to\ell\nu)$+jets:

\begin{align}
\label{eq:illus}
Z_{\rm SR} &\sim  \left( Z_{\rm SR}/Z_{\rm CR}\right)_{\rm MC} \times Z_{\rm CR}^{\rm data}, \nonumber \\ 
Z_{\rm SR} &\sim  \left( Z_{\rm SR}/W_{\rm CR}\right)_{\rm MC} \times W_{\rm CR}^{\rm data},  \\
W_{\rm SR} &\sim \left( W_{\rm SR}/W_{\rm CR}\right)_{\rm MC} \times W_{\rm CR}^{\rm data}. \nonumber
\end{align}

Unique estimates of the $Z(\to\nu\nu)$+jets and $W(\to\ell\nu)$+jets backgrounds in the signal region result from the simultaneous maximum likelihood fit to the control regions and signal region.

The multijet background is estimated from data-driven methods as presented in Section~\ref{sec:dijet}. The data-driven normalizations for the $Z$+jets and $W$+jets backgrounds 
are described in Section~\ref{sec:EW}. The background estimations are validated in control regions with no signal contamination, and are in good agreement with observations in 
the validation control regions, as discussed in Section~\ref{sec:vali}. In SR1 and SR2, the smaller backgrounds of $t\bar{t}$, single top and dibosons are taken from their MC predictions.

Background contributions from the visible Higgs boson decay channels are suppressed by the signal region requirements described in Section~\ref{sec:evt}.

\subsection{Data-driven estimation of the multijet background}
\label{sec:dijet}

Multijet events which have no prompt (from the primary interactions) neutrinos can pass the
$E_{\rm T}^{\rm miss}$ selection due to instrumental effects such as the mis-measurement of the jet energy. Because of the
very large rejection from the $E_{\rm T}^{\rm miss}$ requirement, it is not practical
to simulate this background, so it is estimated using data-driven methods instead.

In the SR2 selections, the multijet background is estimated from data, using a jet smearing method as described in Ref.~\cite{Aad:2012fqa}, 
which relies on the assumption that the $E_{\rm T}^{\rm miss}$ of multijet events is dominated by fluctuations in the detector 
response to jets measured in the data. The estimated multijet background in SR2 is 24 $\pm$ 24 events (a 100\% uncertainty is assigned to the estimate).

In SR1, the multijet background is estimated from data as follows.
 A control region is defined where the $\Delta\phi_{j,E_{\rm T}^{\rm miss}}$ requirement is inverted, so that the $E_{\rm T}^{\rm miss}$ vector is in the
direction of a jet in the event. The resulting sample
is dominated by multijet events.
 The signal region requirements on the leading and sub-leading jet $p_{\rm T}$
and on the $E_{\rm T}^{\rm miss}$ trigger are applied as described in Section~\ref{sec:evt}. 
The efficiency of each subsequent requirement is determined using this sample
and assumed to apply to the signal region with the nominal $\Delta\phi_{j,E_{\rm T}^{\rm miss}}$ requirement.
A systematic uncertainty is assessed based on the accuracy of this assumption in a control region with $|\Delta\eta_{jj}|<3.8$
and in a control region with three jets.
To account for the $\Delta\phi_{j,E_{\rm T}^{\rm miss}}$ requirement itself,
the  $\Delta\phi_{jj}$ requirement is inverted, requiring back-to-back jets
in $\phi$. This sample is also multijet-dominated. Combining all the efficiencies
with the observed control region yield gives an estimate of 2 $\pm$ 2 events  
for the multijet background in SR1.

\subsection{Estimations of the $Z(\rightarrow \nu\nu)$+jets and $W(\rightarrow \ell\nu)$+jets backgrounds}
\label{sec:EW}

To estimate the $Z(\rightarrow \nu\nu)$+jets background, both the $Z(\rightarrow ee/\mu\mu)$+jets and $W(\rightarrow e\nu/\mu\nu)$+jets control regions are employed. 
The $W$+jets background is estimated using $W(\rightarrow e\nu/\mu\nu)$+jets control regions. In the $Z(\rightarrow ee)$+jets control samples for SR1 and $W(\rightarrow\ell\nu)$+jets control samples for SR1 or SR2,  electrons and muons are required to be isolated. Electron isolation is not required in the SR2 $Z(\rightarrow ee)$+jets control sample. For electrons,
the normalized calorimeter isolation transverse energy, i.e. the ratio of the isolation transverse energy to lepton $p_{\rm T}$, is required to be less than 0.28 (0.05) for SR1 (SR2), and the normalized track isolation  is required to be less than 0.1 (0.05) 
within a cone $\Delta R = \sqrt{(\Delta\eta)^2+(\Delta\phi)^2}=0.3$ for SR1 (SR2).
In the SR1 selections, muons must have a normalized calorimeter isolation less than 0.3 (or $<0.18$ if $p_{\mathrm{T}}<25$~GeV) and 
a normalized track isolation less than 0.12 within $\Delta R = 0.3$, whereas in the SR2 selections, the scalar sum of the transverse momentum of tracks in a cone with radius 0.2 around the muon candidate is required to be less than 1.8~GeV.  Electrons and muons are also required to point back to the primary vertex.  The
transverse impact parameter significance must be less than 3$\sigma$ for both the electrons and muons, while
the longitudinal impact parameter must be $<0.4(1.0)$~mm for electrons (muons).

\begin{figure*}[!htbp]
 \begin{center}
\subfloat[]{\label{fig:SR2_met_ZllCRa}\includegraphics[width=0.49\textwidth]{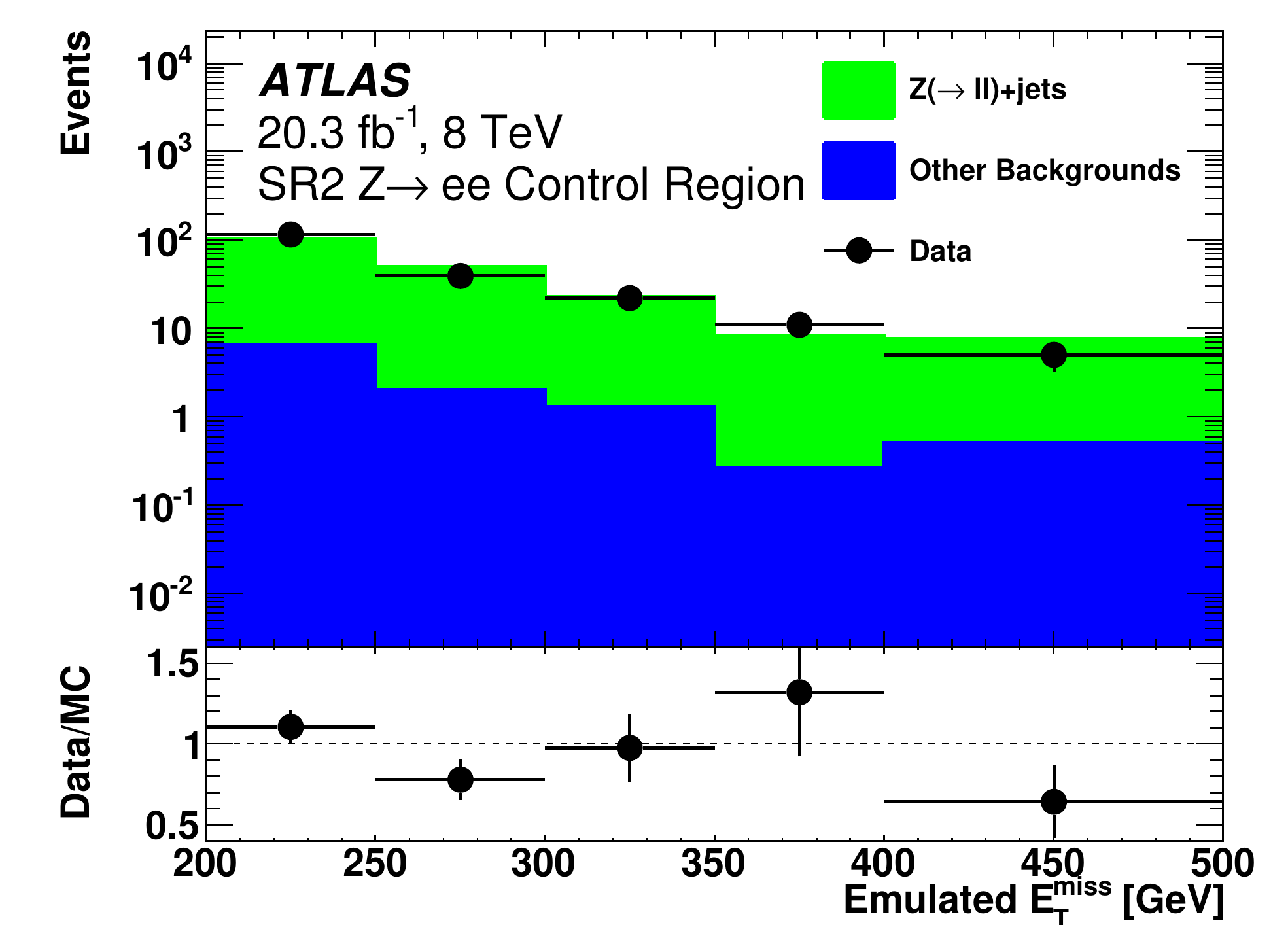}}
\subfloat[]{\label{fig:SR2_met_ZllCRb}\includegraphics[width=0.49\textwidth]{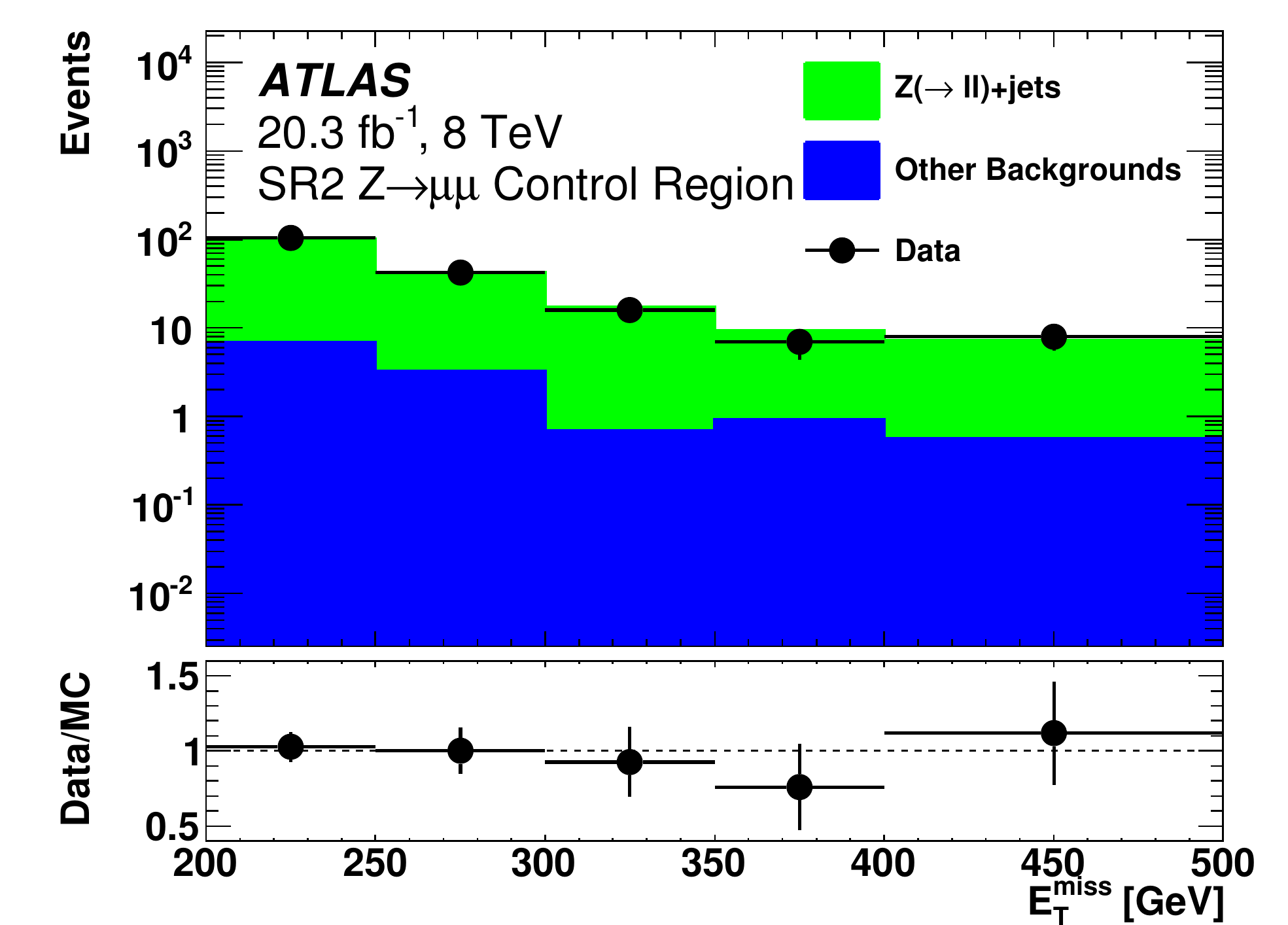}}
  \caption{Data and MC distributions of the $E_{\rm T}^{\rm miss}$ (as described in the text) in the SR2 $Z$+jets control regions (a)~$Z(\rightarrow ee)$+jets and (b)~$Z(\rightarrow \mu\mu)$+jets.\label{fig:SR2_met_ZllCR}}
\end{center}
\end{figure*}

The $Z(\rightarrow ee/\mu\mu)$+jets control regions are defined by selecting
events containing two same-flavour, oppositely charged leptons with $\pT>20$~\GeV~and 
$|m_{\ell\ell}-m_{Z}|<25$~\GeV, where $m_{\ell\ell}$ and $m_Z$ are the dilepton invariant mass and the $Z$-boson mass, respectively.  
In the control sample corresponding to the SR1 selection, the leading lepton
is required to have $\pT>30$~\GeV. 
Triggers requiring a single 
electron or muon with $p_{\rm T}>24$~GeV are used to select the control samples in SR1; in SR2, either a single-electron or $E_{\rm T}^{\rm miss}$ trigger is used.
The inefficiency of the triggers with respect to the offline requirements is negligible.
In order
to emulate the effect of the offline missing transverse momentum selection used in the signal region, the
$E_{\rm T}^{\rm miss}$ quantity is corrected by vectorially adding the electron (SR1 and SR2) and
muon transverse momenta (SR1 only).  
All the $Z(\rightarrow ee/\mu\mu)$+jets events are then required to pass the
 other signal region selections. 
Backgrounds from processes other than $Z(\rightarrow ee/\mu\mu)$+jets are small in these control regions; 
the contributions from non-$Z$ backgrounds are estimated from MC simulation.  For $Z\to ee$ ($Z\to\mu\mu$), the non-$Z$ background is at a level of  1.6\% (0.9\%) of the sample. There is $50$\% uncertainty (mainly due to the limited numbers of MC events) on the non-$Z$ background contamination in the $Z$ control regions. 
The observed yield in the SR1 $Z$ control region, shown in Table~\ref{ZCR_yields}, is larger than the expected yield by 16\% but is
compatible  within the combined statistical uncertainties of MC simulation and data.
\begin{table*}[!htbp]
\begin{center}
\caption{Expected and observed yields for the SR1 $Z(\rightarrow ee/\mu\mu)$+jets control sample in 20.3 fb$^{-1}$ of 2012 data.  Expected contributions
are evaluated using MC simulation, and the uncertainties are statistical only.
\label{ZCR_yields}}
\vspace{6pt}
\begin{tabular}{ l | c c }
\hline\hline
                                & \multicolumn{2}{c}{SR1 $Z$ Control Regions} \\
Background                      & $Z(\rightarrow ee)$+jets     & $Z(\rightarrow\mu\mu)$+jets\\
\hline
QCD $Z\rightarrow\ell\ell$      & 10.4 $\pm$ 1.5       & 14.0 $\pm$ 1.5\\
EW $Z\rightarrow\ell\ell$       &  7.4 $\pm$ 0.8       &  8.2 $\pm$ 0.8\\
Other Backgrounds               &  0.3 $\pm$ 0.2       &  0.2 $\pm$ 0.1\\
\hline
Total                           & 18.1 $\pm$ 1.7       & 22.4 $\pm$ 1.7\\
\hline
Data                            & 22                    & 25 \\
\hline\hline
\end{tabular}
\end{center}
\end{table*}
\begin{table*}[!htbp]
\begin{center}
\caption{Expected and observed yields for the SR2 $Z(\rightarrow ee/\mu\mu)$+jets control sample in 20.3 fb$^{-1}$ of 2012 data.  Expected contributions
are evaluated using MC simulation, and the uncertainties are statistical only.
\label{ZCR_SR2}}
\vspace{6pt}
\begin{tabular}{ l | c c | c c}
\hline\hline
SR2 $Z$ Control Regions & \multicolumn{2}{c}{SR2a}     & \multicolumn{2}{c}{SR2b} \\
Background               & $Z(\rightarrow ee)$+jets   & $Z(\rightarrow\mu\mu)$+jets  & $Z(\rightarrow ee)$+jets   & $Z(\rightarrow\mu\mu)$+jets\\
\hline
QCD $Z\rightarrow\ell\ell$  &  116 $\pm$ 3                 & 121 $\pm$ 4               & 26 $\pm$ 2       &  28 $\pm$ 2\\
EW $Z\rightarrow\ell\ell$   &   17 $\pm$ 1                 &  17 $\pm$ 1               & 16 $\pm$ 1       &  16 $\pm$ 2\\

Other backgrounds           &    8 $\pm$ 1                 &  10 $\pm$ 2               &  2 $\pm$ 1       &   3 $\pm$ 1 \\
\hline
Total                       &  141 $\pm$ 3                 & 148 $\pm$ 5               & 44 $\pm$ 3       &  47 $\pm$ 3\\
\hline
Data                        &  159                         & 139                       & 33               &  38 \\
\hline\hline
\end{tabular}
\end{center}
\end{table*}
In the SR2 control regions, the observed and expected yields differ by 10\% as shown in 
Table~\ref{ZCR_SR2} but are compatible within the total statistical and systematic uncertainties (see Section~\ref{sec:syst}). 
\begin{figure*}[!htbp]
\begin{center}
\includegraphics[width=0.75\columnwidth]{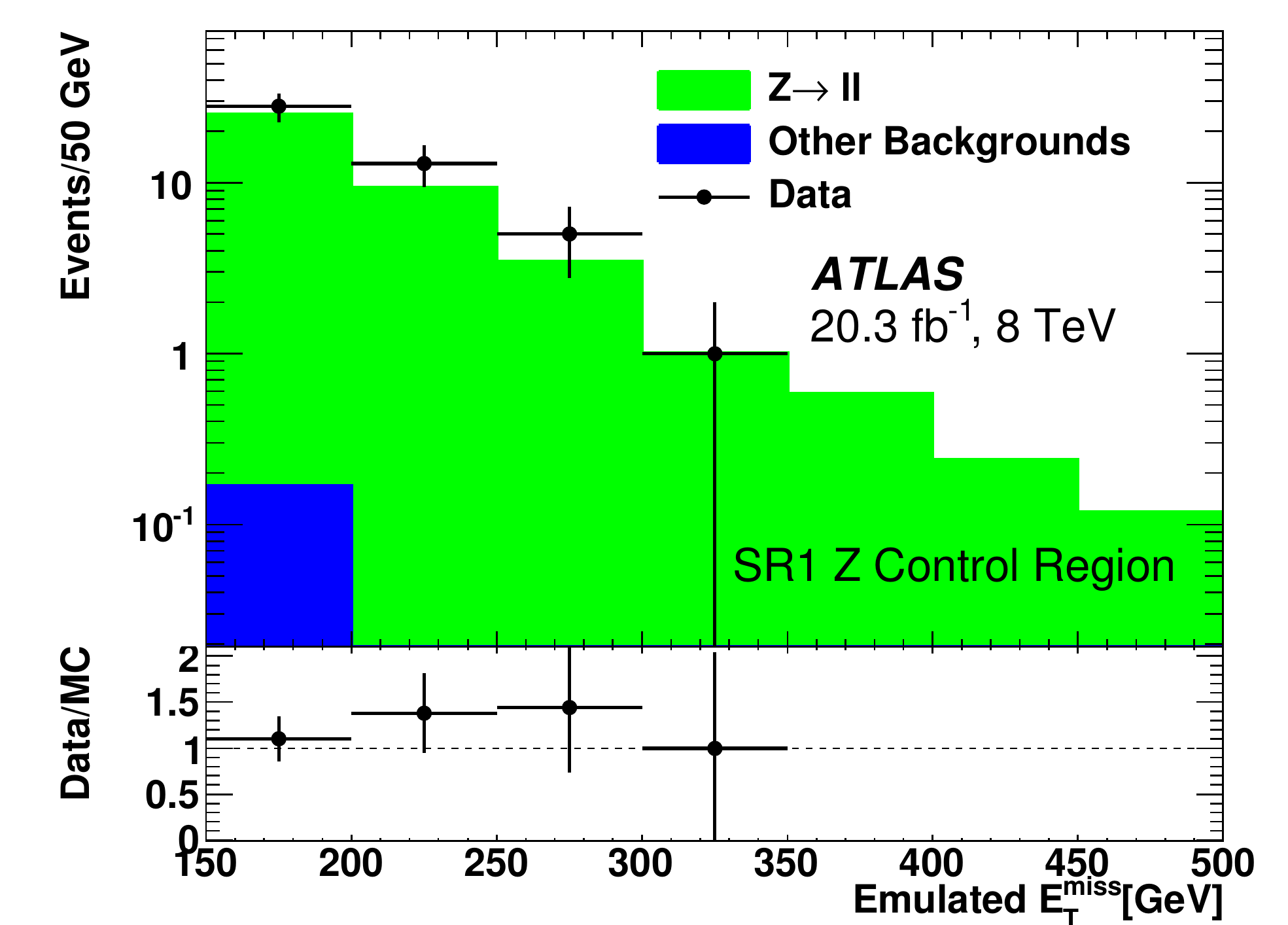}
  \caption{Data and MC distributions of the emulated $E_{\rm T}^{\rm miss}$ (as described in the text) in the SR1 $Z(\rightarrow ee/\mu\mu)$+jets 
control region.\label{fig:SR1_met_ZllCR}}
\end{center}
\end{figure*}
The emulated $E_{\rm T}^{\rm miss}$ distributions for the $Z$ control regions are shown in Figures~\ref{fig:SR1_met_ZllCR} and~\ref{fig:SR2_met_ZllCR} for SR1 and SR2 respectively. Because the muon momentum is excluded from the $E_{\rm T}^{\rm miss}$ definition in SR2, the "emulated" label is omitted from Figure~\ref{fig:SR2_met_ZllCRb}.

The $W(\rightarrow e\nu/\mu\nu)$+jets control regions are similarly defined by
selecting events containing one lepton with transverse momentum $p_{\rm T}>30$~\GeV~(25~\GeV)~in the case of SR1 (SR2), and no additional leptons with $p_{\rm T}>20$~\GeV.
The $E_{\rm T}^{\rm miss}$ is emulated in the same way as for the $Z\rightarrow ee/\mu\mu$ control region and events are required to pass the signal region
selections on jets and $E_{\rm T}^{\rm miss}$. In SR1, the contributions of
the three lepton flavours to the total $W\rightarrow\ell\nu$ background after all the requirements are 20\% for $W\rightarrow e\nu$,
 20\% for $W\rightarrow\mu\nu$, and 60\% for $W\rightarrow\tau\nu$.
The equal contributions from $W\rightarrow e\nu$ and $W\rightarrow\mu\nu$ suggest that these are events where the lepton is below its $p_{\rm T}$ threshold
or sufficiently far forward to escape the jet veto, and not events where the lepton is misidentified as a tag jet (since muons deposit little energy in the calorimeter and
would therefore not be identified as a jet).  This expectation is checked explicitly for the case of $W\rightarrow\tau\nu$, by using MC truth information
about the direction of the $\tau$-lepton to find the $\Delta R$ between the $\tau$-lepton and the nearest reconstructed tag jet.  The component with $\Delta R_{j,\tau}<0.4$
is  completely negligible after the signal region requirements, indicating that the lepton tends to be recoiling against the tag jets rather than being aligned with them.  For SR1, four $W$ control regions are considered using different charge samples for $W^+/W^- \to e\nu/\mu \nu$ since $W(\to \ell\nu)$+jets is not charge symmetric as shown in Table~\ref{wlnu_requirementflow_8TeV_nomT}, whereas in
 SR2, only two control regions $W(\rightarrow e\nu/\mu\nu)$+jets are used as shown in Table~\ref{WCR_SR2}.
\begin{table*}[!htbp]
\begin{center}
\caption{Expected and observed yields for the SR1 $W\rightarrow\ell\nu$ control sample, after all requirements in 20.3 fb$^{-1}$ of 2012 data.  The multijet background is estimated using
the data-driven method described in the text; all other contributions are evaluated using MC simulation.  Only the statistical uncertainties are shown.
\label{wlnu_requirementflow_8TeV_nomT}}
\vspace{6pt}
\begin{tabular}{ l | c c c c }
\hline
\hline
                                & \multicolumn{4}{c}{SR1 $W$ Control Regions} \\
Background                      & $W^{+}\rightarrow e\nu$       & $W^{-}\rightarrow e\nu$       & $W^{+}\rightarrow\mu\nu$      & $W^{-}\rightarrow\mu\nu$\\
\hline
QCD $W\rightarrow\ell\nu$                         & 92.3 $\pm$ 7.2            & 55.1 $\pm$ 5.3            & 85.5 $\pm$ 7.0        & 43.8 $\pm$ 4.6\\
EW $W\rightarrow\ell\nu$                          & 99.4 $\pm$ 4.0            & 52.5 $\pm$ 2.9            & 81.9 $\pm$ 3.7        & 39.1 $\pm$ 2.5\\
QCD $Z\rightarrow\ell\ell$      &  3.4 $\pm$ 0.6            &  4.4 $\pm$ 0.9            &  6.4 $\pm$ 1.1        &  5.0 $\pm$ 0.9\\
EW $Z\rightarrow\ell\ell$       &  2.5 $\pm$ 0.3            &  2.9 $\pm$ 0.3            &  2.7 $\pm$ 0.3        &  3.2 $\pm$ 0.3\\
Multijet                        & 28.0 $\pm$ 6.8            & 28.0 $\pm$ 6.8            &  1.6 $\pm$ 2.6        &  1.6 $\pm$ 2.6\\
Other backgrounds               &  4.0 $\pm$ 0.7            &  1.8 $\pm$ 0.4            &  3.2 $\pm$ 0.7        &  1.0 $\pm$ 0.3\\
\hline
Total                           & 230 $\pm$ 11              & 145 $\pm$ 9               & 181 $\pm$ 8         & 93.7 $\pm$ 5.9\\
\hline
Data                            & 225                       & 141                       & 182                   & 98 \\
\hline
\hline
\end{tabular}
\end{center}
\end{table*}
\begin{table*}[!htbp]
\begin{center}
\caption{Expected and observed yields for the SR2 $W(\rightarrow e\nu/\mu\nu)$+jets control sample in 20.3 fb$^{-1}$ of 2012 data.  Expected contributions
are evaluated using MC simulation, and the uncertainties are statistical only. The discrepancy in the $W(\rightarrow \mu\nu)$+jets SR2b control region is
due to a mis-modelling of the $W$ $p_{\rm T}$. The agreement improves when the systematic uncertainties (discussed in Section~\ref{sec:syst}) are included. 
\vspace{6pt}
\label{WCR_SR2}}
\begin{tabular}{ l | c c | c c}
\hline\hline
\multicolumn{1}{c}{SR2 $W$ Control Regions} & \multicolumn{2}{c}{SR2a}                           & \multicolumn{2}{c}{SR2b} \\
Background               & $W(\rightarrow e\nu)$+jets   & $W(\rightarrow\mu\nu)$+jets  & $W(\rightarrow e\nu)$+jets   & $W(\rightarrow\mu\nu)$+jets\\
\hline
QCD $W\rightarrow\ell\nu$   &  595 $\pm$ 12                & 906 $\pm$ 15              & 122 $\pm$ 5       &  201 $\pm$ 7\\
EW  $W\rightarrow\ell\nu$   &  149 $\pm$ 5                 & 214 $\pm$ 6               & 121 $\pm$ 4       &  184 $\pm$ 5\\
QCD $Z\to\ell\ell$          &   5.8 $\pm$ 0.9              & 23 $\pm$   1.6            &  1.6 $\pm$ 0.4    &   4.5 $\pm$ 0.6        \\
EW  $Z\to\ell\ell$          &   0.4 $\pm$ 0.1              & 0.5 $\pm$ 0.2              &  2.0 $\pm$ 0.4       &  2.7 $\pm$ 0.8        \\
Multijet                    &   13 $\pm$ 3                 &  0 $\pm$ 0                &  3 $\pm$ 1        &   0 $\pm$ 0        \\ 
Other backgrounds           &    44 $\pm$ 4                &  78 $\pm$ 7               &  13 $\pm$ 2       &   19 $\pm$ 3 \\
\hline
Total                       &  807 $\pm$ 14                &  1222 $\pm$ 18            & 263 $\pm$ 7       &  411 $\pm$ 9\\
\hline
Data                        &  783                         & 1209                      & 224               & 295 \\
\hline\hline
\end{tabular}
\end{center}
\end{table*}

\begin{figure*}[!tbph]
 \centering
\subfloat[]{\includegraphics[width=0.49\textwidth]{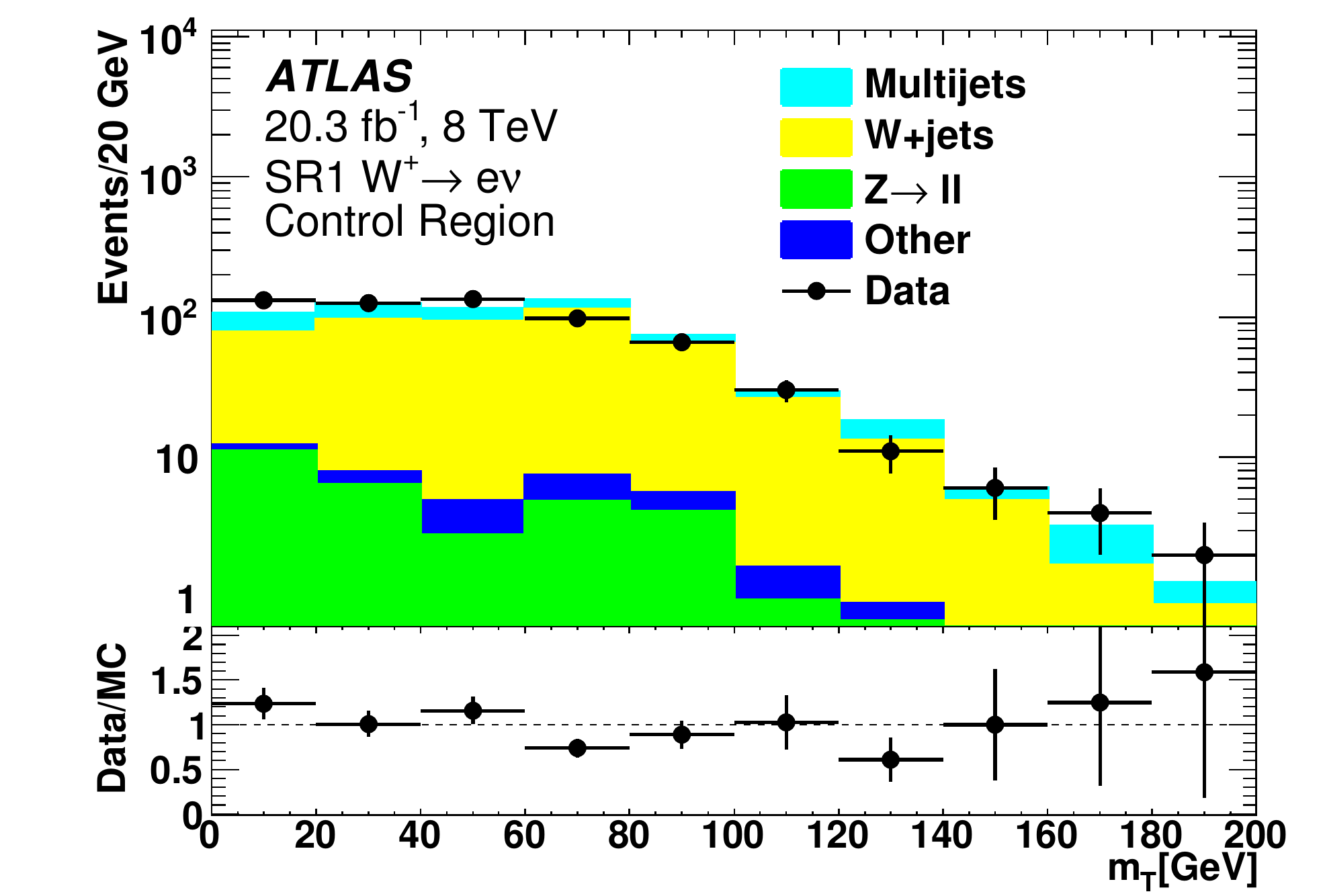}}
\subfloat[]{\includegraphics[width=0.49\textwidth]{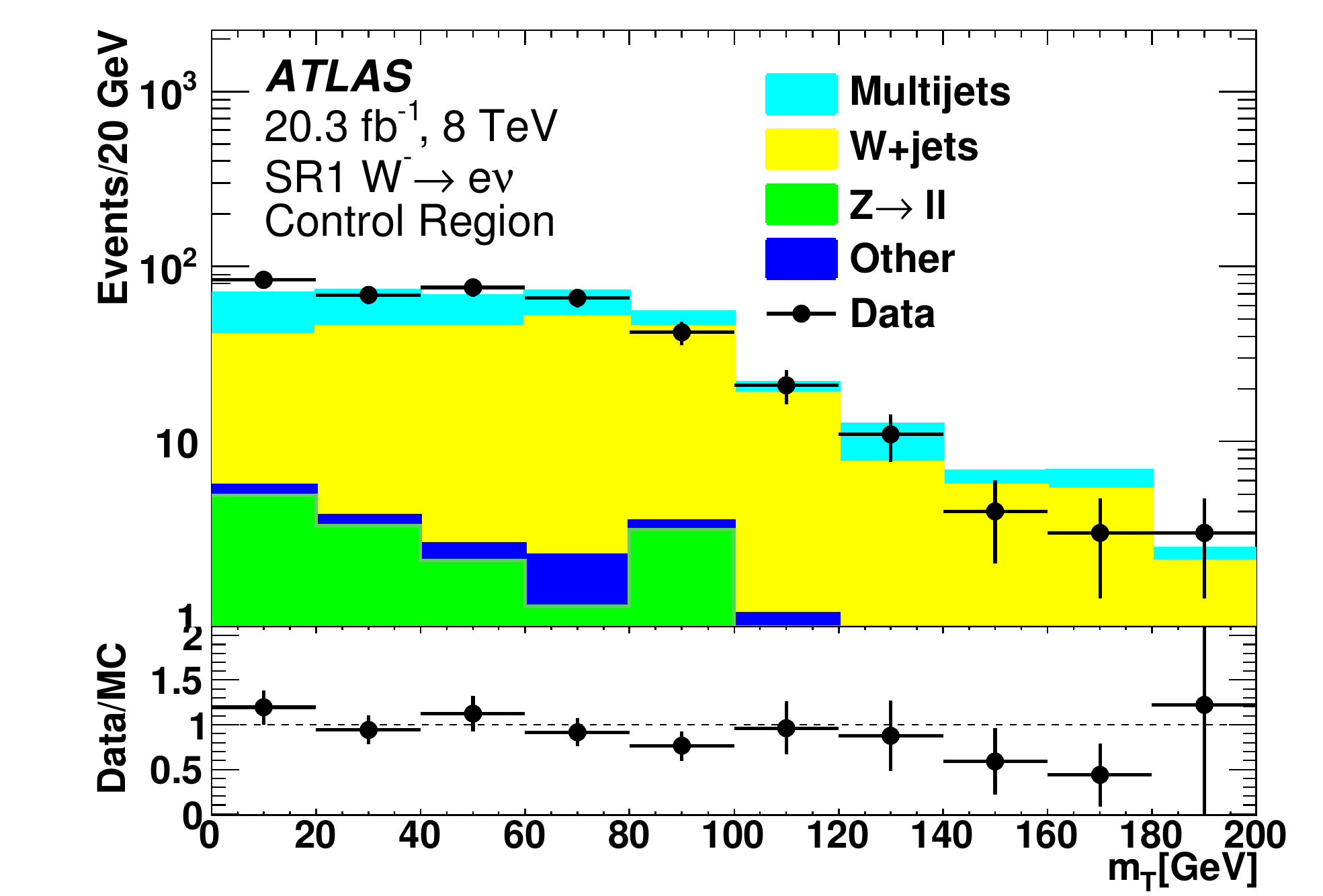}} \\
\subfloat[]{\includegraphics[width=0.49\textwidth]{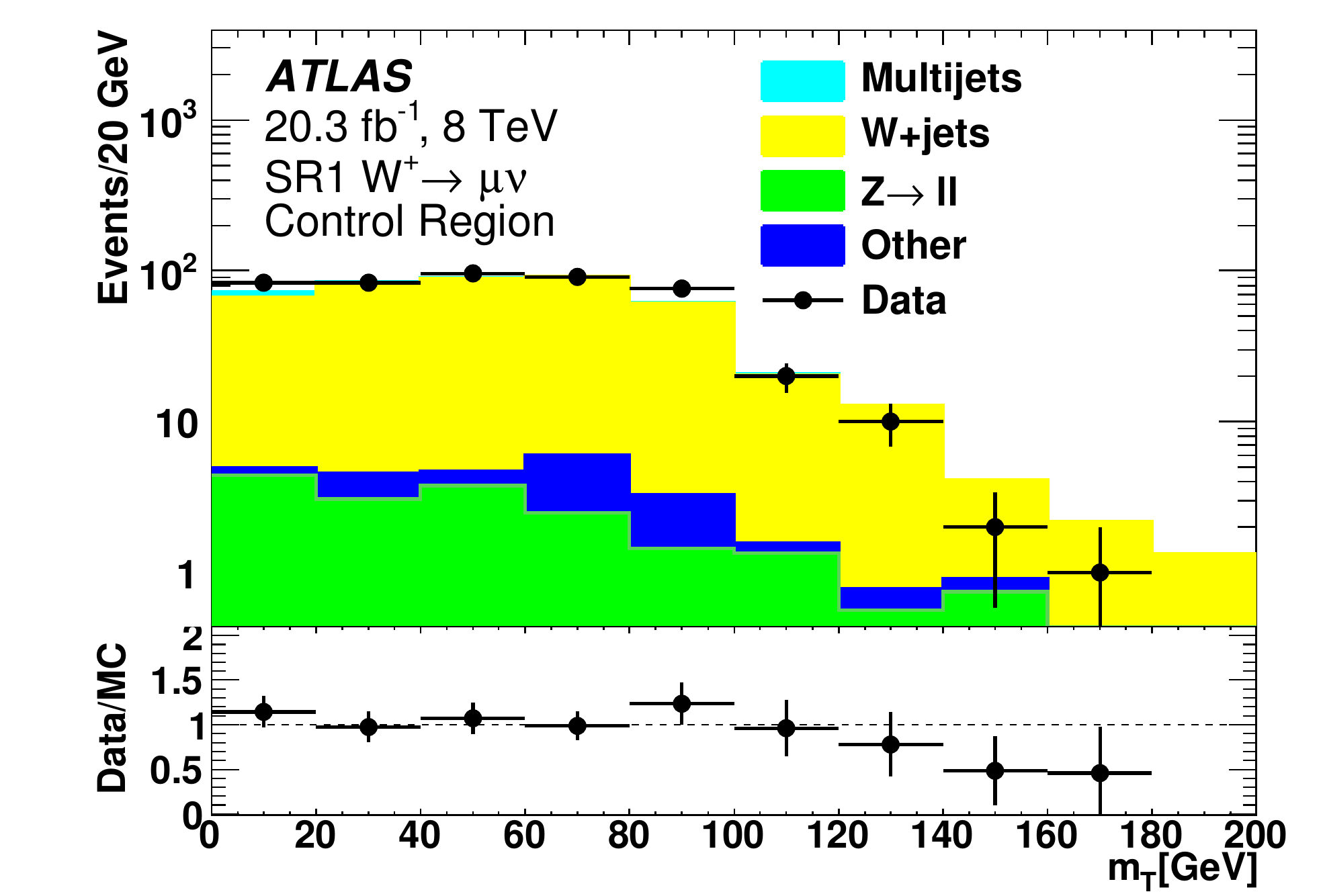}}
\subfloat[]{\includegraphics[width=0.49\textwidth]{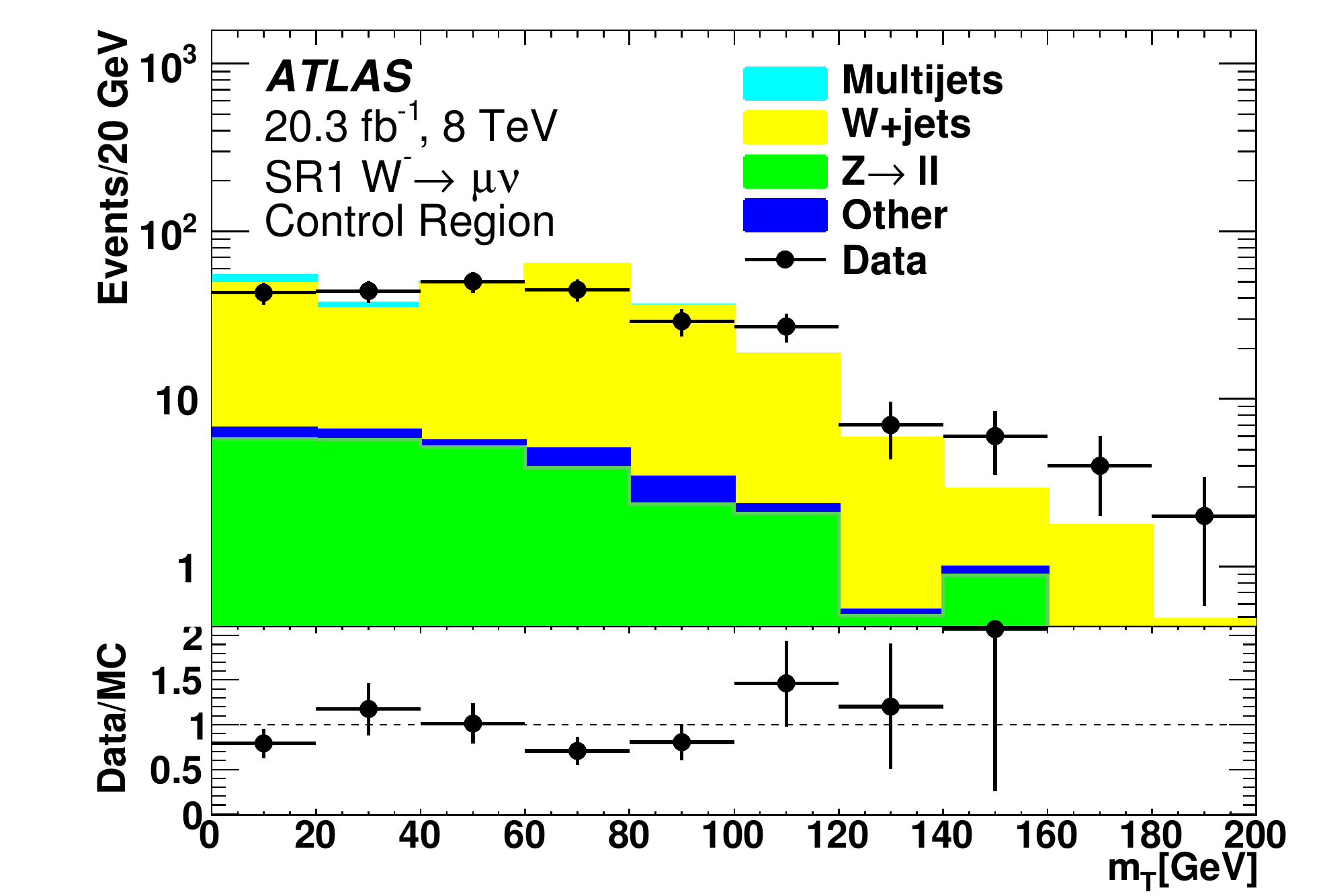}}

  \caption{The transverse mass distributions used in the SR1 $W$+jets control regions after all requirements  except for the $E_{\rm T}^{\rm miss}>150$~GeV 
requirement: (a) $W^+\rightarrow e^+\nu$, (b) $W^-\rightarrow e^-\nu$, (c) $W^+\rightarrow\mu^+\nu$ and (d) $W^-\rightarrow\mu^-\nu$.
  \label{mT_fit_plots_allrequirements}}
\end{figure*}

In the $W(\rightarrow e\nu/\mu\nu)$+jets control regions corresponding to the SR1 selection, a fit to the transverse mass defined in Eq.~(\ref{eq:transmass}) is used to estimate the multijet background. In order to obtain an explicit measurement and uncertainty for the background
from multijets, no requirements are made on $E_{\rm T}^{\rm miss}$ and $m_{\rm T}$. Because the multijet background does not have a prompt neutrino, the $E_{\rm T}^{\rm miss}$ tends
to be lower and to point in the direction of the jet that was  
misidentified as a lepton. As a result,  the multijet background tends 
to have significantly lower $m_{\rm T}$ than the $W$+jets contribution.
Control samples modelling the jets misidentified as leptons in
multijet events are constructed by selecting events that pass
 the $W$+jets control region selection, except for certain lepton
identification criteria:  for electrons, some of the EM calorimeter shower
shape requirements are loosened and fully identified electrons are
removed, while for muons, the transverse impact parameter ($d_{0}$) requirement which
suppresses muons originating from heavy-flavour jets is reversed. 
 To obtain the normalization of the multijet background in the $W$+jets control region,
templates of the $m_{\rm T}$ distribution for
processes with prompt leptons are taken from MC simulation.    %
Shape templates for the backgrounds from multijet events are constructed
by summing the observed yields in control samples obtained by inverting the lepton identification and $d_0$ requirements, and
subtracting the expected contributions from $W$+jets and $Z$+jets events using MC.  %
Since the misidentified-jet samples are expected to be charge-symmetric, 
the same shape template and normalization factor
is used to model both charge categories of a given lepton flavour ($e$ or $\mu$).
To determine the $W(\rightarrow\ell\nu)$+jets background normalization, a fit to the transverse mass $m_{\rm T}$ of the lepton and $E_{\rm T}^{\rm miss}$ is used. 
The $W(\rightarrow\ell\nu)$+jets contribution, however, is not charge symmetric,
so the different charge samples are kept separate in the simultaneous fit to four
$m_{\rm T}$ distributions, one for each lepton flavour and charge combination shown in Figure~\ref{mT_fit_plots_allrequirements}.
There are three free normalizations in the fit: one for events with a prompt lepton, one for events where a jet is misidentified as
an electron, and one for events where a jet is misidentified as a muon.
The normalization factor for the prompt leptons in the $m_{\rm T}$
fit is 0.95 $\pm$ 0.05 (stat). 

\begin{figure*}[!tbph]
 \centering
\subfloat[]{\includegraphics[width=0.49\textwidth]{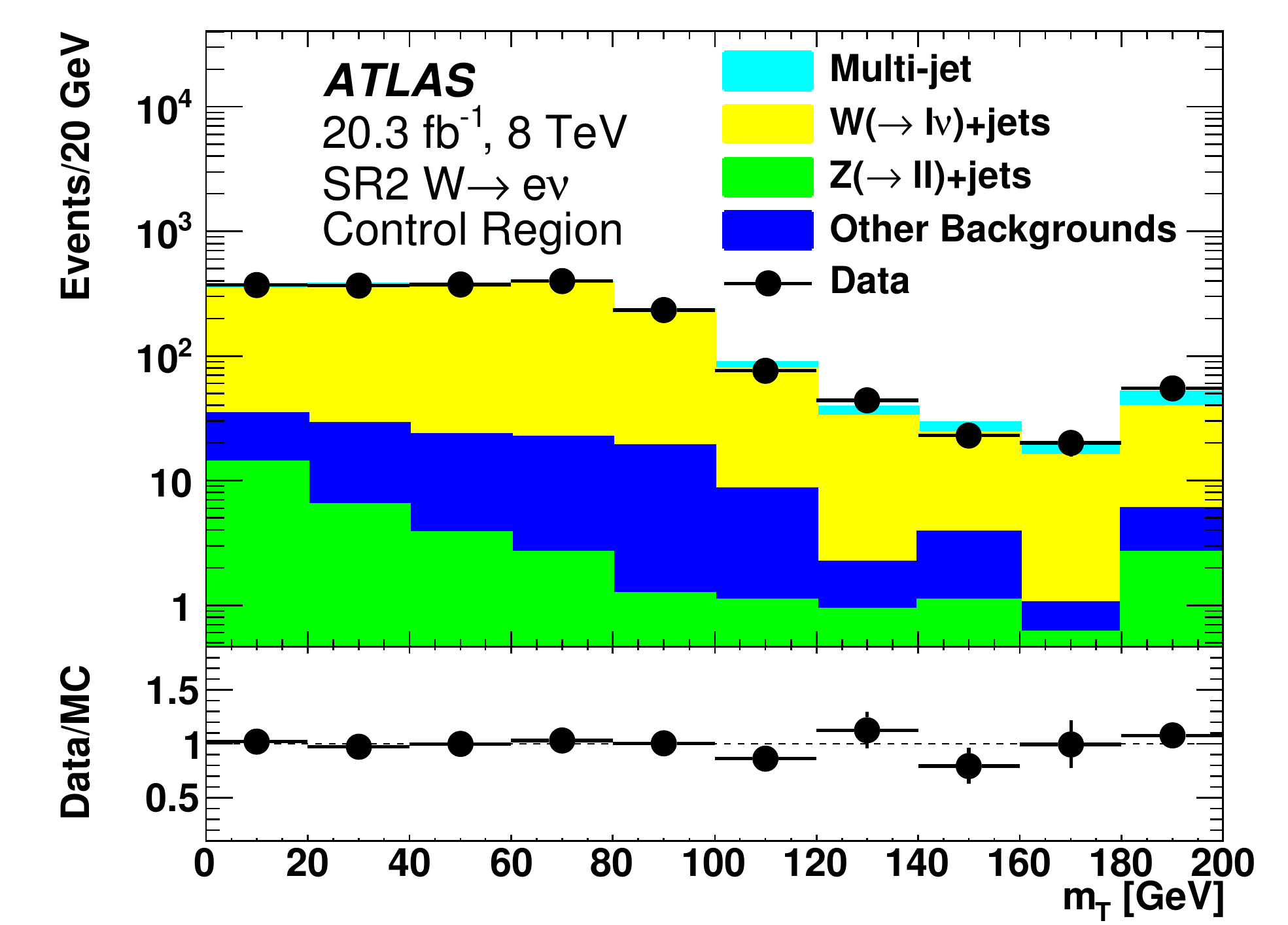}}
\subfloat[]{\includegraphics[width=0.49\textwidth]{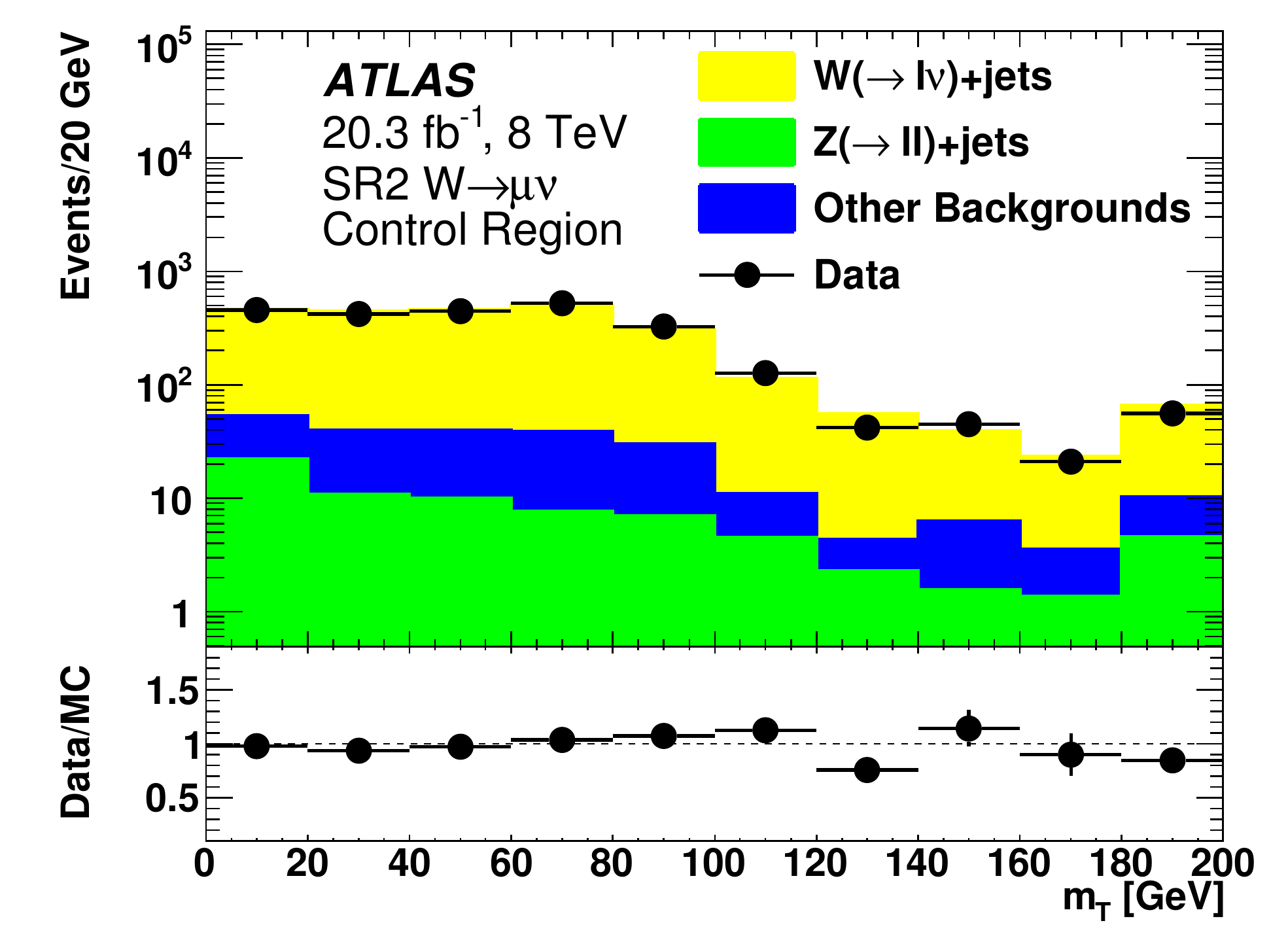}}

  \caption{The transverse mass distributions in the SR2 $W$+jets control regions after all requirements: (a) $W\rightarrow e\nu$ and (b) $W\rightarrow \mu\nu$.
  \label{sr2_mT_fit_plots_allrequirements}}
\end{figure*}

In the $W\rightarrow e\nu$ control region corresponding to the SR2 selections,
the background from multijet events 
 is rejected by requiring that the $E_{\rm T}^{\rm miss}$ (corrected by vectorially adding the electron transverse momentum) be larger than 25~\GeV~and that the
transverse mass be in the range $40<m_{\rm T}<100$~GeV. The selected electron is required to pass both the track and calorimeter isolation requirements. The tight requirements on electron isolation and $E_{\rm T}^{\rm miss}$ greatly reduce the multijet background relative to the other backgrounds. The residual multijet background in the $W\to e\nu$ control region is at the level of 1\% of the total control region background, with an uncertainty of 100\%. For the $W\to \mu\nu$ control region corresponding to the SR2 selections, the selected muon is required to pass only the track 
isolation requirement and the transverse mass is required to be in the range $30<m_{\rm T}<100$~GeV. An attempt is made to estimate the residual multijet background in the $W\to \mu\nu$ control region using a control sample with inverted muon isolation. The residual background from multijet events is negligible. 
 Figure~\ref{sr2_mT_fit_plots_allrequirements} shows the $m_{\rm T}$ distributions in the SR2 $W$+jets control regions.

\subsection{Validation of Data-Driven Estimations}
\label{sec:vali}

 To validate the background estimates for SR1,
two signal-depleted neighbouring regions are defined
by %
(1)
reversing the veto against three-jet events and requiring that the third jet in the event has transverse momentum $p_{\rm
  T}^{j3}>40$~GeV, and (2) reversing both the jet veto with a $p_{\rm T}^{j3}>30$~GeV requirement and the jet rapidity gap with a $|\Delta\eta_{jj}|<3.8$ requirement.  Good agreement between expectation and
observation is found in these validation regions, as shown in Table~\ref{VR_yields}.

\begin{table*}[!htbp]
\begin{center}
\caption{Expected and observed yields for the validation regions in
  20.3 fb$^{-1}$ of data. 3-jet: reversal of the veto against three-jet events by requiring $p_{\rm
  T}^{j3}>40$~\GeV; and 3-jet and $|\Delta\eta_{jj}|<3.8$: requirements of  both $|\Delta\eta_{jj}|<3.8$ and
$p_{\rm T}^{j3}>30$~\GeV.  Contributions from $W$+jets and $Z$+jets are
  normalized to data-driven estimates. The $W$+jets and $Z$+jets
  uncertainties include MC statistics from both the selected region
  and the corresponding control region, and the number of data events
  in the control regions. The other numbers are evaluated using MC simulation and their uncertainties indicate only statistical
  uncertainty.
\label{VR_yields}}
\vspace{6pt}
\begin{tabular}{ l | c c  }
\hline\hline
Process                         & 3-jet                     & 3-jet and $|\Delta\eta_{jj}|<3.8$\\
\hline
ggF signal                      & $6.2 \pm 3.1$                                 & -\\
VBF signal                      & $19.9 \pm 1.4$                   & $4.7 \pm 0.6$\\
\hline
$Z(\rightarrow\nu\nu$)+jets       & $97 \pm 10$                     & $111 \pm 10$\\
$W(\rightarrow\ell\nu$)+jets      & $78.5 \pm 6.5$                      & $73 \pm 10$\\
Multijet                         & $19.9 \pm 21.8$                                  & -\\
Other backgrounds               & $2.2 \pm 0.3$                      & $0.5 \pm 0.1$\\
\hline
Total                           & $198 \pm 25$                       & $185 \pm 14$\\
\hline
Data                            & 212                                &195\\

\hline\hline
\end{tabular}
\end{center}
\end{table*}

\section{Systematic uncertainties}
\label{sec:syst}
The experimental uncertainties on the MC predictions for signals 
and backgrounds are dominated by uncertainties in the jet energy
scale and resolution~\cite{atlaspaper:JES7TeV}. This includes effects such as the $\eta$ dependence of the energy
scale calibration and the dependence of the energy response on the 
jet flavour composition, where flavour refers to the gluon or light quark initiating the jet.  
Uncertainties related to the lepton identification in the control
regions and lepton vetoes are negligible. Luminosity uncertainties~\cite{Aad:2013ucp} are applied to the signal and background yields that
are obtained from MC simulation. 

Theoretical uncertainties on the $W$+jets and $Z$+jets contributions to both the
signal and control regions are assessed using {\sc Sherpa}, and cross-checked with  
{\sc MCFM}~\cite{Campbell:2011bn} and {\sc VBFNLO}~\cite{Arnold:2008rz} for the EW and QCD processes 
respectively, and by a comparison between {\sc Sherpa} and {\sc Alpgen}~\cite{ALPGEN} for the latter process.  In all cases, the 
uncertainties are determined by independently varying the factorization and
renormalization scales by factors of 2 and $1/2$, keeping their ratio within 0.5--2.0. The parton distribution function uncertainties are evaluated with the CT10
error sets \cite{ct10}. The uncertainty on the ggF yield due to the jet selection is evaluated using Stewart--Tackmann method~\cite{Stewart:2011cf}. Uncertainty in the $p_{\mathrm{T}}$ distribution of the Higgs boson in ggF is evaluated from scale variations in {\sc HRes} following the re-weighting of the $p_{\mathrm{T}}$ distribution~\cite{deFlorian:2011xf,Grazzini:2013mca} as mentioned in Section~\ref{sec:intro}.  To assess the level of theoretical uncertainty on the jet
veto, the variation in the predicted VBF cross section with respect to shifts in the renormalization and factorization scales as well as with respect to uncertainty in the parton-shower modelling is measured using {\sc Powheg-box} NLO generator matched to {\sc Pythia} and to {\sc Herwig}. The effect of the parton shower on the QCD $W$+jets and $Z$+jets background estimations is obtained by comparing simulated samples with different
parton shower models. As shown in Table~\ref{tab:syst}, where the main systematic uncertainties are summarized, using the MC predictions of $Z_{\rm SR}/W_{\rm CR}$ and $W_{\rm SR}/W_{\rm CR}$ ratios reduces the systematic uncertainties in the final $Z$+jets and $W$+jets background estimates. The $Z(\to\ell\ell)$+jets$/W(\to\ell\nu$)+jets ratio is checked in data and MC, and no discrepancy larger than 10\% is observed, consistent with the residual theory uncertainties on the $Z_{\rm SR}/W_{\rm CR}$ ratios shown in Table~\ref{tab:syst}.

\begin{table*}[!htbp]
\begin{center}
\caption{Detector and theory uncertainties (\%) after all SR or CR selections. For each source of uncertainty, where relevant,
the first and second rows correspond to the uncertainties in SR1 and SR2 respectively. The ranges of uncertainties in the $Z$ or $W$ column correspond to uncertainties in the $Z$+jets and $W$+jets MC yields in the SR or CR. The
search uses the uncertainties in the ratios of SR to CR yields shown in the last column. 
\label{tab:syst}}
\vspace{6pt}
\begin{tabular}{l|c|c|c|c}
\hline\hline
Uncertainty        & VBF                  & ggF                  & $Z$ or $W$                             & $Z_{\rm SR}/W_{\rm CR}$ or $W_{\rm SR}/W_{\rm CR}$                    \\
\hline
\multirow{2}{*}{Jet energy scale}                & 16                   & 43                   &  17--33                                     & 3--5                     \\
                   & 9                   & 12                   &   0--11                                     & 1--4                     \\
\hline
\multirow{2}{*}{Jet energy resolution}  & Negligible               & Negligible                   &  Negligible                                        & Negligible                          \\
                   & 3.1                  & 3.2                  &   0.2--7.6                                  & 0.5--5.8                         \\
\hline
\multirow{2}{*}{Luminosity}         & \multirow{2}{*}{2.8} & \multirow{2}{*}{2.8} & \multirow{2}{*}{2.8}                        &   \multirow{2}{*}{Irrelevant}                        \\
                   &                      &                      &                                             &                          \\
\hline
\multirow{2}{*}{QCD scale}          & \multirow{2}{*}{0.2} & \multirow{2}{*}{7.8} & 5--36 & 7.8--12                     \\
                                    &                      &                      & 7.5--21                  & 1--2                     \\
\hline
\multirow{2}{*}{PDF}                &     2.3 & \multirow{2}{*}{7.5} & 3--5                     & \multirow{2}{*}{1--2}     \\
                                    &     2.8 &                      & 0.1-2.6                  &                          \\
\hline
\multirow{2}{*}{Parton shower}      &    \multirow{4}{*}{4.4}      &                      & \multirow{2}{*}{9--10}           & \multirow{2}{*}{5}     \\
                                    &                              &     41                 &                                             &             \\
\cline{1-1}\cline{4-5}
\multirow{2}{*}{Veto on third jet}   &                              &  29                   &   \multirow{2}{*}{Negligible}       & \multirow{2}{*}{Negligible}        \\
                                    &                              &                 &                                             &      \\
\hline
\multirow{2}{*}{Higgs boson $p_{\rm T}$}  &  \multirow{2}{*}{Negligible}     & \multirow{2}{*}{9.7}                  &         \multirow{2}{*}{Irrelevant}                                     &      \multirow{2}{*}{Irrelevant}                     \\
                   &                      &                   &                                             &                          \\
\cline{1-1}\cline{3-5}

\hline
\multirow{2}{*}{MC statistics}      &   2     &  46         & 2.3--6.4 &  \multirow{2}{*}{3.3--6.6} \\
                  &                  0.6      &     13      & 0.8--4.5 &                          \\
\hline\hline
\end{tabular}
\end{center}
\end{table*}

\section{Results}
\label{sec:results}

\begin{figure*}[!htbp]
\begin{center}
  \subfloat[$E_{\rm T}^{\rm miss}$ distribution]{
\includegraphics[width=0.5\columnwidth]{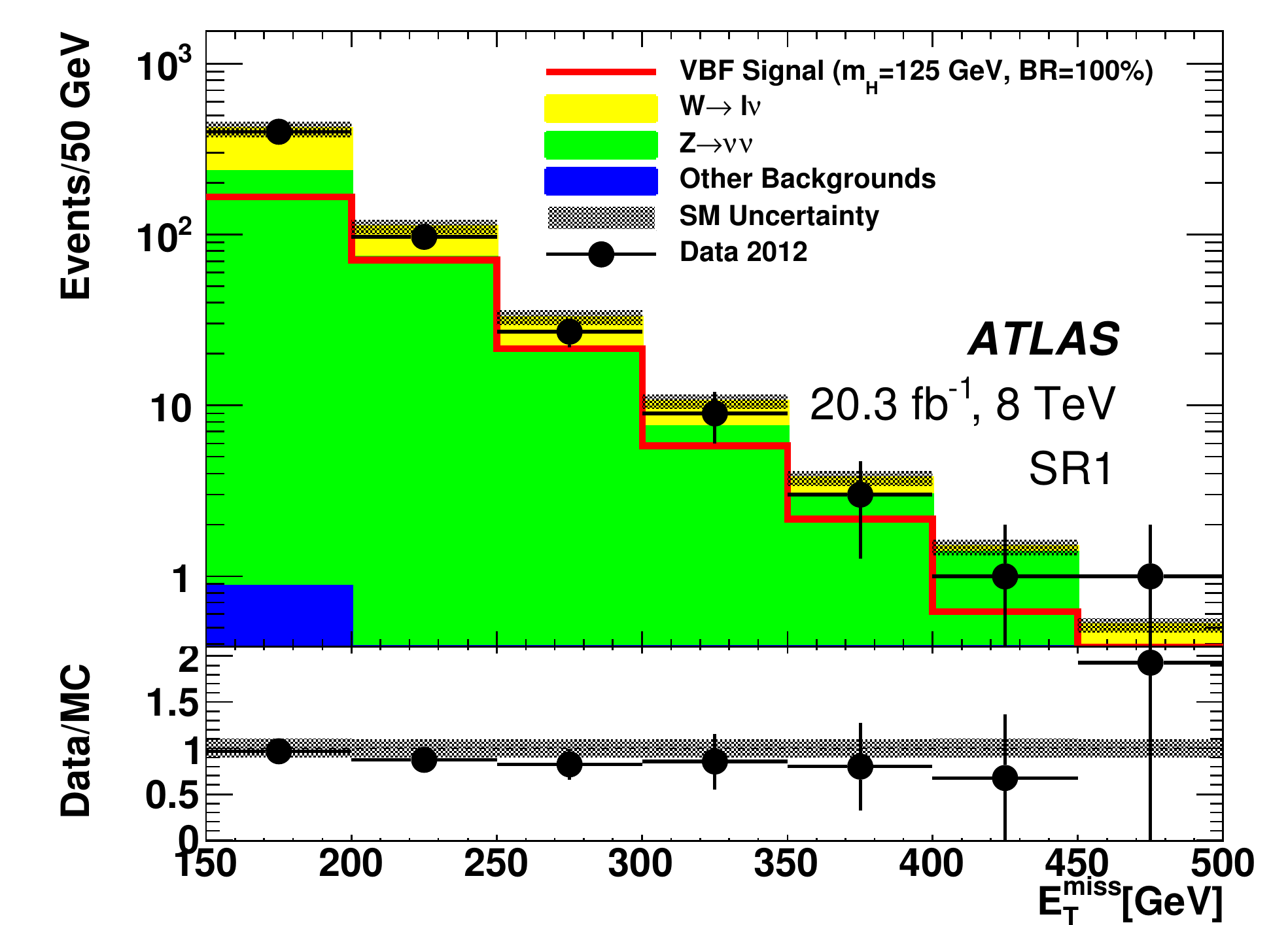}\label{met_distribution}}
  \subfloat[$m_{jj}$ distribution]{
\includegraphics[width=0.5\columnwidth]{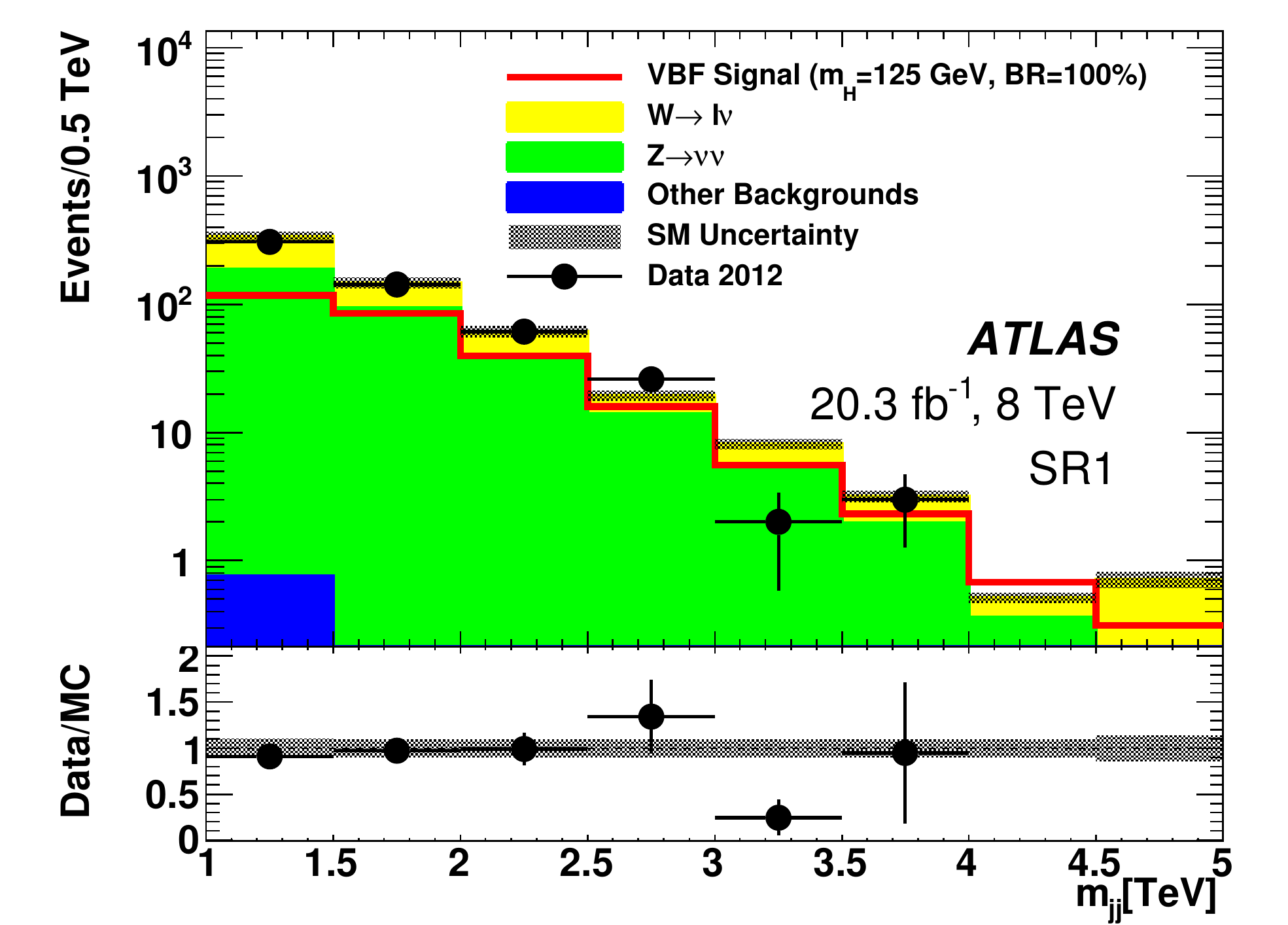}\label{mjj_distribution}}
\end{center}
  \caption{Data and MC distributions after all the requirements in SR1 for (a) $E_{\rm T}^{\rm miss}$ and (b) the dijet invariant mass $m_{jj}$.  The background 
histograms are normalized to
the values in Table~\ref{requirementflow_SR_8TeV}.  The VBF signal (red histogram) is normalized
  to the SM VBF Higgs boson production cross section with BF($H\rightarrow \mathrm{invisible}$) = 100\%.
  \label{Distributions}
}
\end{figure*}

\begin{figure*}[!htbp]
\begin{center}
  \subfloat[$E_{\rm T}^{\rm miss}$ distribution]{
\includegraphics[width=0.5\columnwidth]{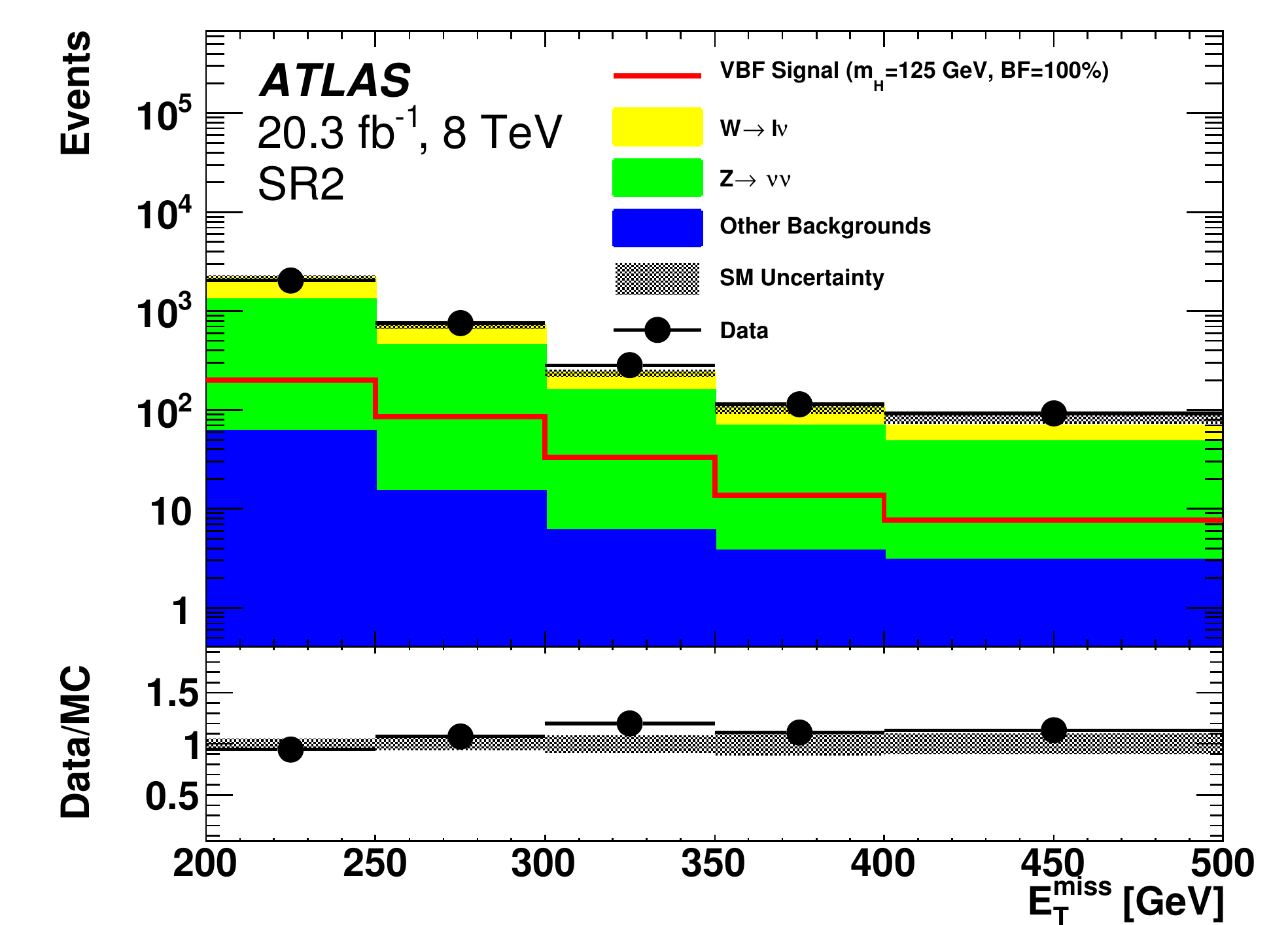}\label{sr2_met_distribution}}
  \subfloat[$m_{jj}$ distribution]{
\includegraphics[width=0.5\columnwidth]{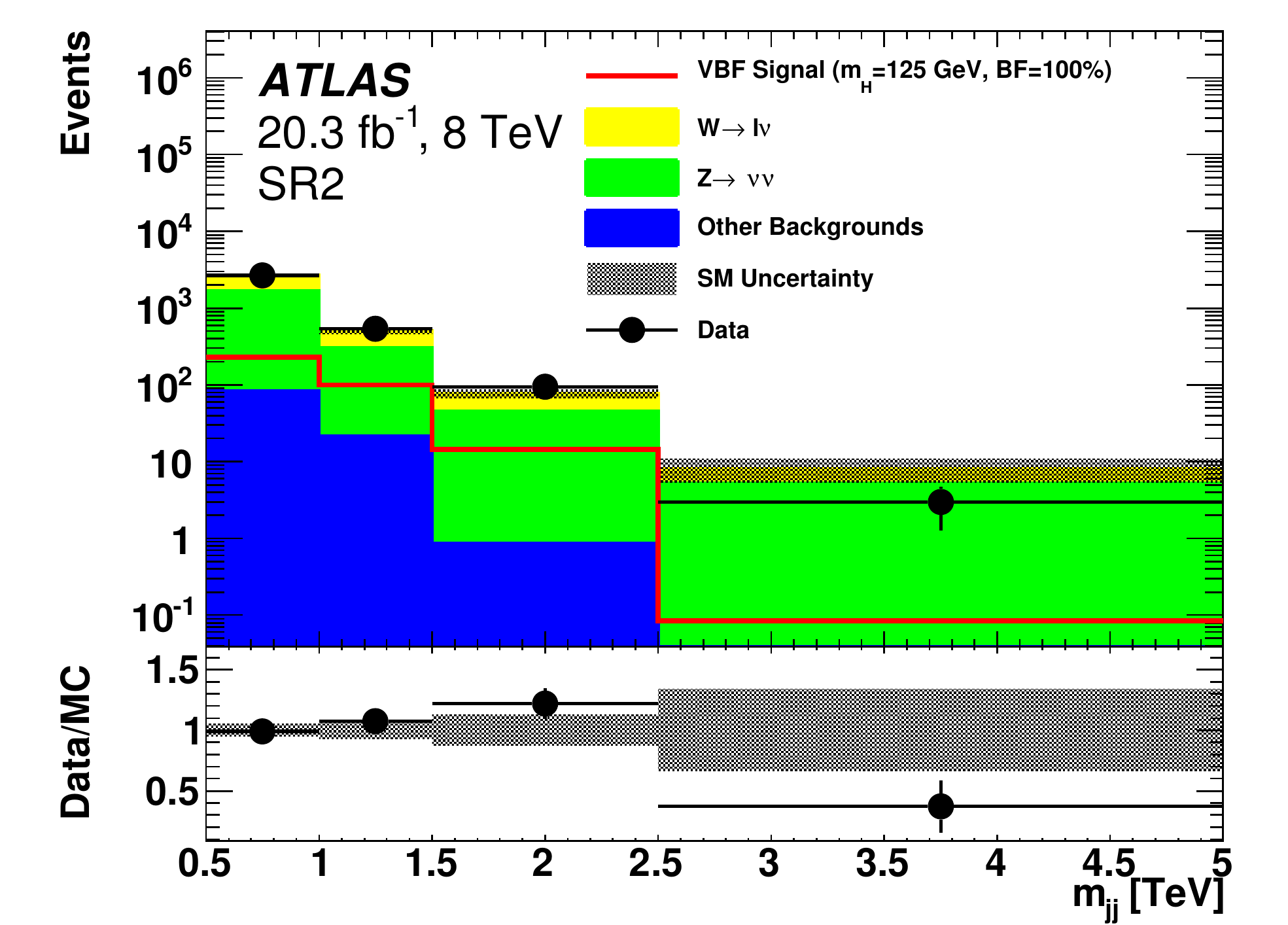}\label{sr2_mjj_distribution}}

\end{center}
  \caption{Data and MC distributions after all the requirements in SR2 for (a) $E_{\rm T}^{\rm miss}$ and (b) the dijet invariant mass $m_{jj}$. The background histograms are 
normalized to the values in Table~\ref{requirementflow_SR_8TeV}.
 The VBF signal is normalized to the SM VBF Higgs boson production cross section with BF($H\rightarrow \mathrm{invisible}$) = 100\%.
  \label{orth_mjj}}
\end{figure*}

Figures~\ref{Distributions} and~\ref{orth_mjj}  show the $E_{\rm T}^{\rm miss}$ and the  $m_{jj}$ distributions after imposing the requirements of SR1 and SR2 respectively. There is good agreement between the data and the background expectations from the SM, and no statistically significant excess is observed in data.
The limit on the branching fraction of $H\rightarrow$ invisible is computed using
a maximum-likelihood fit to the yields in the signal regions and the
$W(\rightarrow e\nu/\mu\nu)$+jets and $Z(\rightarrow ee/\mu\mu)$+jets control samples following
the CL$_{S}$  modified frequentist formalism~\cite{CLs}
with a profile likelihood-ratio test statistic~\cite{2011EPJC...71.1554C}.  
Expected signal and background distributions in the signal and control regions are determined from
MC predictions, with the exception of the multijet backgrounds, which use the data-driven methods described in Section~\ref{sec:CR}.
Systematic uncertainties are parameterized as Gaussian constrained nuisance parameters.
The nuisance parameter for each individual source of uncertainty is shared among
the expected yields so that its correlated effect is taken into account.
 The relative weight of the $Z(\rightarrow ee/\mu\mu)$+jets and $W(\rightarrow e\nu/\mu\nu)$+jets in the  control regions
is determined by the maximization of the likelihood function. 

One global likelihood function including all three signal regions and the six corresponding control regions
is constructed with only the signal yields and correlated uncertainties coupling the search regions.
The theoretical uncertainties are taken to be uncorrelated between the EW and QCD processes and uncorrelated with
the scale uncertainty on the signal. The uncertainties which are treated as correlated between the regions are:
\begin{itemize}
\item Uncertainty in the luminosity measurements. This impacts  the predicted rates of the signals and the backgrounds that are estimated using MC simulation, namely ggF and VBF signals, and $t\bar{t}$, single top, and diboson backgrounds.
\item Uncertainties in the absolute scale and resolution of the reconstructed jet energy.
\item Uncertainties in the modelling of the parton shower.
\item Uncertainties in renormalization and factorization scales.
\end{itemize}
Table~\ref{requirementflow_SR_8TeV} shows signal, background and data events after the global fit including the effects of systematic uncertainties, 
MC statistical uncertainties in the control and signal regions, and the data statistical uncertainties in the control regions. The post-fit values of the 
$Z$+jets and $W$+jets background normalization scale factors $k_i$, discussed in Section~\ref{sec:CR}, are $0.95\pm 0.21$, $0.87 \pm 0.17$ and $0.74 \pm 0.12$ for SR1, SR2a and SR2b and their control regions, respectively. As shown in 
Table~\ref{requirementflow_SR_8TeV}, the signal-to-background ratio is 0.53 in SR1, and  0.09 and 0.21 in SR2a and SR2b respectively, for BF($H\to$~invisible) $=$ 100\%. 
\begin{table*}[tbh]
\begin{center}
\caption{
Estimates of the expected yields and their total uncertainties for SR1 and SR2 in 20.3 fb$^{-1}$ of 2012 data.
The $Z(\rightarrow\nu\nu)$+jets, $W(\rightarrow\ell\nu)$+jets, and multijet background estimates are data-driven.
The other backgrounds and the ggF and VBF signals are determined from MC simulation.
The expected signal yields are shown for  $m_{H}=125$~\GeV~and are normalized to BF($H\to$~invisible) $=$ 100\%.
The $W$+jets and $Z$+jets statistical uncertainties result from the number of MC events in each signal and
corresponding control region, and from the
number of data events in the control region. 
}
\vspace{6pt}
\begin{tabular}{ l | r@{$\pm$} c  | r@{$\pm$} c | r@{$\pm$} c } %

\hline
\hline
\textbf{Signal region} & \multicolumn{2}{c|}{SR1} & \multicolumn{2}{c|}{SR2a} & \multicolumn{2}{c}{SR2b}\\
Process                         &  \multicolumn{2}{c|}{}  & \multicolumn{2}{c|}{}  & \multicolumn{2}{c}{}   \\
\hline
ggF signal                      &  20   &  15    & 58     & 22    & 19    & 8  \\
VBF signal                      &  286  &  57    & 182    & 19    & 105   & 15 \\
\hline
$Z(\rightarrow\nu\nu)$+jets     &  339  &  37    & 1580   & 90    & 335   & 23 \\
$W(\rightarrow\ell\nu)$+jets    &  235  &  42    & 1010   & 50    & 225   & 16 \\
Multijet                        &  2    &  2     & 20     & 20    & 4     & 4  \\
Other backgrounds               &  1    &  0.4   & 64     & 9     & 19    & 6  \\
\hline
Total background                & 577  &  62    & 2680   & 130   & 583   & 34 \\
\hline
Data                            & \multicolumn{2}{c|}{539} & \multicolumn{2}{c|}{2654} & \multicolumn{2}{c}{636} \\
\hline
\hline
\end{tabular}
\label{requirementflow_SR_8TeV}
\end{center}
\end{table*}
Fits to the likelihood function are performed separately for each signal region and their combination, and the 95\% CL limits on
BF($H\rightarrow \mathrm{invisible}$) are shown in Table~\ref{tab:comb}.
\begin{table*}[!hbpt]
\begin{center}
\caption{Summary of limits on BF($H\rightarrow$ invisible) for 20.3~fb$^{-1}$ of 8~TeV data in the individual search regions and
their combination, assuming the SM cross section for $m_H=125$~GeV.
\vspace{6pt}
\label{tab:comb}}
\begin{tabular}{ l | c c c c c | c }
\hline
\hline
Results               & Expected      & $+1\sigma$    & $-1\sigma$    & $+2\sigma$    & $-2\sigma$    & Observed\\
\hline
\hline
SR1                   & 0.35          & 0.49          & 0.25          & 0.67          & 0.19          & 0.30\\
SR2                   & 0.60          & 0.85          & 0.43          & 1.18          & 0.32          & 0.83\\
Combined Results      & 0.31          & 0.44          & 0.23          & 0.60          & 0.17          & 0.28\\
\hline
\hline
\end{tabular}
\end{center}
\end{table*}

The agreement between the data and the background expectations in SR1 is also expressed as a model-independent 95\% CL upper limit on the fiducial cross
section
\begin{eqnarray}
\sigma_{\rm fid} & = & \sigma \times {\rm BF} \times {\cal A}, \\
                 & = & \frac{N}{ {\cal L} \times \epsilon}, 
\end{eqnarray}
where the acceptance ${\cal A}$ is the fraction of events within the fiducal phase space defined at the MC truth level using the SR1 selections in Section~\ref{sec:evt}, $N$ the accepted number of events, ${\cal L}$ the integrated luminosity and $\epsilon$ the selection efficiency defined as the ratio of selected events to those in the fiducial phase space. Only the systematic uncertainties on the backgrounds and the integrated luminosity are taken into account in the upper limit on $\sigma_{\rm fid}$, shown in Table~\ref{tab:mu}.
\begin{table*}[!hbpt]
\begin{center}
\caption{Model-independent 95\% CL upper limit on the fiducial cross section for non-SM processes $\sigma_{\rm fid}$ in SR1.
\vspace{6pt}
\label{tab:mu}}
\begin{tabular}{ l | c c c c c | c }
\hline
\hline
SR1                          & Expected      & $+1\sigma$    & $-1\sigma$    & $+2\sigma$    & $-2\sigma$    & Observed \\
\hline
\hline
Fiducial cross section [fb]  & 4.78          & 6.32          & 3.51          & 8.43          & 2.53          & 3.93\\
\hline
\hline
\end{tabular}
\end{center}
\end{table*}
In SR1, the acceptance and the event selection efficiency, estimated from simulated VBF $H\to ZZ \to 4\nu$ events, are $(0.89 \pm 0.04)$\% and $(94 \pm 15)$\% respectively. The uncertainties have been divided such that the theory uncertainties are assigned to the acceptance and the experiment uncertainties are assigned to the efficiency.

\section{Model interpretation}
\label{sec:interpretation}

In the Higgs-portal dark-matter scenario, a dark sector is coupled to the Standard Model via
the Higgs boson~\cite{Kanemura:2010sh,Djouadi:2011aa} by introducing a WIMP dark-matter singlet $\chi$ that only 
couples to the SM Higgs doublet. In this model, assuming that the dark-matter particle is lighter than half the Higgs boson mass, one would search for Higgs boson decays to undetected (invisible) dark-matter particles, e.g. $H\to\chi\chi$.
 The upper limits on the branching fraction to invisible particles directly determine 
the maximum allowed decay width to the invisible particles
\begin{equation}
\Gamma_H^{\rm inv}=\frac{\textrm{BF}(H\to\, \textrm{invisible)}}{1-\textrm{BF}(H\to\, \textrm{invisible)}}\times\Gamma_H,
\end{equation}
where $\Gamma_H$ is the SM decay width of the Higgs boson. 
Adopting the formulas from Ref.~\cite{Djouadi:2011aa}, the decay width of the
Higgs boson to the invisible particles can be written as
\begin{eqnarray}
\Gamma^\textrm{inv}_{H\rightarrow SS} &=&
\frac{ \lambda_{HSS}^2 v^2 \beta_S }{ 64 \pi m_H }, \\
\Gamma^\textrm{inv}_{H\rightarrow VV} &=&
\frac{ \lambda_{HVV}^2 v^2 m_H^3 \beta_V }{ 256 \pi m_V^4 }
\Bigg(1-4\frac{m_V^2}{m_H^2}+12\frac{m_V^4}{m_H^4}\Bigg), \\
\Gamma^\textrm{inv}_{H\rightarrow ff} &=&
\frac{ \lambda_{Hff}^2 v^2 m_H \beta_f^3 }{ 32 \pi \Lambda^2 },
\end{eqnarray}
for the scalar, vector and Majorana-fermion dark matter, respectively.
The parameters $\lambda_{HSS}$, $\lambda_{HVV}$, $\lambda_{Hff}/\Lambda$ are the
corresponding coupling constants, $v$ is the vacuum expectation value of the SM Higgs doublet, $\beta_\chi=\sqrt{1-4m_\chi^2/m_H^2}$ ($\chi$ $=$ $S$, $V$, $f$), and $m_\chi$ is the WIMP mass. In the Higgs-portal model, the Higgs boson is assumed to be the only mediator in the WIMP--nucleon scattering, and the
WIMP--nucleon cross section can be written in a general spin-independent form.  Inserting the couplings and masses for each spin scenario gives:
\begin{eqnarray}
\sigma^{\rm SI}_{SN} &=&
\frac{ \lambda_{HSS}^2 }{ 16 \pi m_H^4 }
\frac{ m_N^4 f_N^2 }{ (m_S+m_N)^2 }, \\
\sigma^{\rm SI}_{VN} &=&
\frac{ \lambda_{HVV}^2 }{ 16 \pi m_H^4 }
\frac{ m_N^4 f_N^2 }{ (m_V+m_N)^2 }, \\
\sigma^{\rm SI}_{fN} &=&
\frac{ \lambda_{Hff}^2 }{ 4 \pi \Lambda^2 m_H^4 }
\frac{ m_N^4 m_f^2 f_N^2 }{ (m_f+m_N)^2 },
\end{eqnarray}
where $m_N$ is the nucleon mass, and $f_N$ is the form factor associated to the Higgs boson--nucleon coupling and computed using lattice QCD~\cite{Djouadi:2011aa}. The numerical values for all the parameters in the equations above
are given in Table~\ref{tab:portal_param}.
\begin{table*}[!htbp]
\caption{Parameters in the Higgs-portal dark-matter model.
\label{tab:portal_param}}
\vspace{6pt}
\centering
\begin{tabular}{l|c|c}
\hline
\hline \vspace{-11pt} \\
Vacuum expectation value & $v/\sqrt{2}$ & $174\,\GeV$ \\
Higgs boson mass & $m_H$ & $125\,\GeV$ \\
Higgs boson width & $\Gamma_H$ & $4.07\,\MeV$ \\
Nucleon mass & $m_N$ & $939\,\MeV$ \\
Higgs--nucleon coupling form factor & $f_N$ & $0.33^{+0.30}_{-0.07}$ \vspace{5pt}\\
\hline\hline
\end{tabular}
\end{table*}

\begin{figure*}[!htbp]
\centering
\includegraphics[width=0.75\columnwidth]{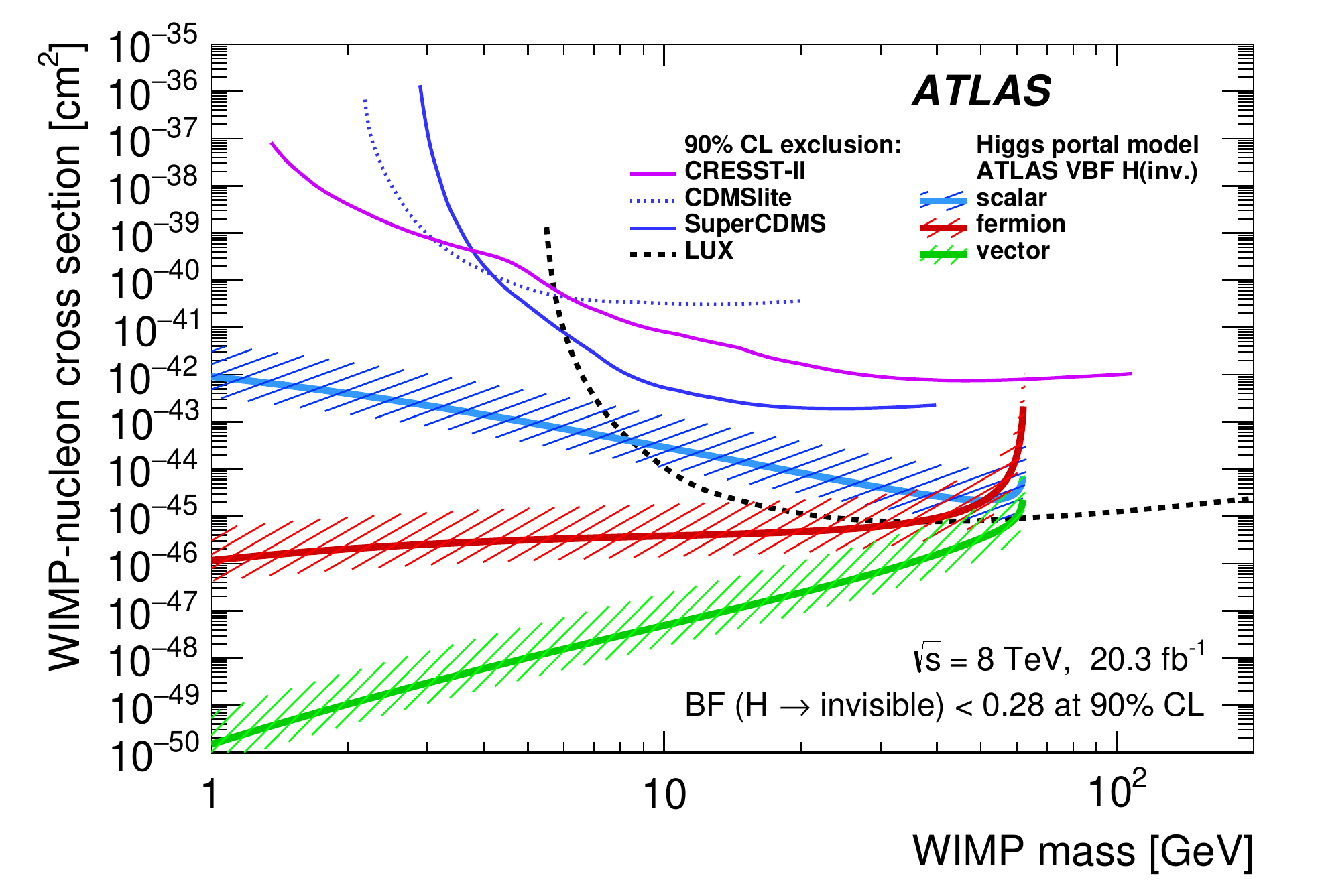}
\caption{
The WIMP--nucleon cross section as a function of the WIMP mass.
The exclusion limits~\cite{Angloher:2014myn,Agnese:2013jaa,Agnese:2014aze,Aprile:2013doa,Akerib:2013tjd} 
of the direct detection experiments are compared to the ATLAS results from the BF($H\rightarrow$~invisible) limit in the Higgs-portal scenario,
translated into the WIMP--nucleon cross section using the formulas from Ref.~\cite{Djouadi:2011aa}.
The exclusion limits are shown at 90\% CL. The error bands on the ATLAS results indicate the uncertainty coming from the different estimations of the Higgs--nucleon coupling form factor~\cite{Young:2009zb,Toussaint:2009pz}.
\label{higgs_portal_dark_matter}}
\end{figure*}

The inferred 90\% CL branching fraction limit for $H\to$~invisible, translated into an upper bound on the scattering cross section between nucleons and WIMP, is shown in Figure~\ref{higgs_portal_dark_matter} compared to the results from direct detection experiments. The WIMP--nucleon cross-section limits resulting from searches for invisible Higgs boson decays extend
from low WIMP mass to half the Higgs boson
mass, and are complementary to the results provided by direct detection experiments that have limited sensitivity to WIMP with mass of the order of 10~GeV and lower~\cite{Angloher:2014myn, 
Akerib:2013tjd,Agnese:2013rvf,Agnese:2014aze,Aalseth:2014jpa,Bernabei:2008yi,Aprile:2013doa}. 
This is expected as
the LHC has no limitations for the production of low-mass particles, whereas the
recoil energies produced in the interactions of sub-relativistic
WIMP with nuclei in the apparatus of a direct detection experiment are
often below the sensitivity threshold for small WIMP masses.
The aforementioned correlation between the branching fraction of Higgs boson decays to invisible particles and the WIMP--nucleon cross section is presented in the effective field theory framework, assuming that the new physics scale is ${\cal O}({\rm a\, few})$\,TeV, well above the scale probed at SM Higgs boson mass. Adding a renormalizable mechanism for generating the fermion and vector WIMP masses could modify the correlation between the WIMP--nucleon cross section and the branching fraction of Higgs  boson decays to invisible particles~\cite{Baek:2014jga}.

\section{Conclusions}
\label{sec:conc}

A search for Higgs boson decays to invisible particles is presented. The search uses data events with two forward jets and large missing transverse momentum, 
collected with the ATLAS detector from 20.3 fb$^{-1}$ of $pp$ collisions at $\sqrt{s}=8$~TeV at the LHC. Assuming the SM production cross section, acceptance and efficiency for invisible decays of a Higgs boson with a mass of 125~GeV, a 95\% CL upper
bound is set on the BF($H \to$~invisible) at 0.28. The results are interpreted in the Higgs-portal dark-matter model where the 90\% CL limit on the BF($H \to$~invisible) is converted into upper bounds on the dark--matter nucleon scattering cross section as a function of the dark-matter particle mass. The ATLAS limits are complementary to the results from the direct dark-matter detection experiments.

\section*{Acknowledgments}

We thank CERN for the very successful operation of the LHC, as well as the
support staff from our institutions without whom ATLAS could not be
operated efficiently.

We acknowledge the support of ANPCyT, Argentina; YerPhI, Armenia; ARC,
Australia; BMWF, Austria; AHAS, Azerbaijan; SSTC, Belarus; CNPq and FAPESP,
Brazil; NSERC, NRC and CFI, Canada; CERN; CONICYT, Chile; CAS, MOST and
NSFC, China; COLCIENCIAS, Colombia; MSMT CR, MPO CR and VSC CR, Czech
Republic; DNRF, DNSRC and Lundbeck Foundation, Denmark; ARTEMIS, European
Union; IN2P3-CNRS, CEA-DSM/IRFU, France; GNAS, Georgia; BMBF, DFG, HGF, MPG
and AvH Foundation, Germany; GSRT, Greece; ISF, MINERVA, GIF, DIP and
Benoziyo Center, Israel; INFN, Italy; MEXT and JSPS, Japan; CNRST, Morocco;
FOM and NWO, Netherlands; RCN, Norway; MNiSW, Poland; GRICES and FCT,
Portugal; MERYS (MECTS), Romania; MES of Russia and ROSATOM, Russian
Federation; JINR; MSTD, Serbia; MSSR, Slovakia; ARRS and MVZT, Slovenia;
DST/NRF, South Africa; MICINN, Spain; SRC and Wallenberg Foundation,
Sweden; SER, SNSF and Cantons of Bern and Geneva, Switzerland; NSC, Taiwan;
TAEK, Turkey; STFC, the Royal Society and Leverhulme Trust, United Kingdom;
DOE and NSF, United States of America.

The crucial computing support from all WLCG partners is acknowledged
gratefully, in particular from CERN and the ATLAS Tier-1 facilities at
TRIUMF (Canada), NDGF (Denmark, Norway, Sweden), CC-IN2P3 (France),
KIT/GridKA (Germany), INFN-CNAF (Italy), NL-T1 (Netherlands), PIC (Spain),
ASGC (Taiwan), RAL (UK) and BNL (USA) and in the Tier-2 facilities
worldwide.

\clearpage
\printbibliography

\clearpage
\input{atlas_authlist}

\end{document}

%% file: atlas_authlist.tex
\begin{flushleft}
{\Large The ATLAS Collaboration}

\bigskip

G.~Aad$^{\rm 85}$,
B.~Abbott$^{\rm 113}$,
J.~Abdallah$^{\rm 151}$,
O.~Abdinov$^{\rm 11}$,
R.~Aben$^{\rm 107}$,
M.~Abolins$^{\rm 90}$,
O.S.~AbouZeid$^{\rm 158}$,
H.~Abramowicz$^{\rm 153}$,
H.~Abreu$^{\rm 152}$,
R.~Abreu$^{\rm 116}$
% ...
\bigskip
\\
% ...
$^{11}$ Institute of Physics, Azerbaijan Academy of Sciences, Baku, Azerbaijan\\
% ...
$^{85}$ CPPM, Aix-Marseille Universit{\'e} and CNRS/IN2P3, Marseille, France\\
% ...
$^{90}$ Department of Physics and Astronomy, Michigan State University, East Lansing MI, United States of America\\
% ...
$^{107}$ Nikhef National Institute for Subatomic Physics and University of Amsterdam, Amsterdam, Netherlands\\
% ...
$^{113}$ Homer L. Dodge Department of Physics and Astronomy, University of Oklahoma, Norman OK, United States of America\\
% ...
$^{116}$ Center for High Energy Physics, University of Oregon, Eugene OR, United States of America\\
% ...
$^{151}$ Institute of Physics, Academia Sinica, Taipei, Taiwan\\
$^{152}$ Department of Physics, Technion: Israel Institute of Technology, Haifa, Israel\\
$^{153}$ Raymond and Beverly Sackler School of Physics and Astronomy, Tel Aviv University, Tel Aviv, Israel\\
% ...
$^{158}$ Department of Physics, University of Toronto, Toronto ON, Canada\\
% ...
\end{flushleft}